\documentclass[fleqn,usenatbib]{mnras}          
\usepackage{newtxtext,newtxmath}                
\usepackage{newtxmath}
\usepackage[T1]{fontenc}

\DeclareRobustCommand{\VAN}[3]{#2}
\let\VANthebibliography\thebibliography
\def\thebibliography{\DeclareRobustCommand{\VAN}[3]{##3}\VANthebibliography}

\usepackage{graphicx}	                        
\usepackage{amsmath}	                        
\usepackage{verbatim}                           
\usepackage{caption}

\title[Massive, quiescent galaxies at high-redshift]{On the origin of
star-formation quenching in massive galaxies at $z \gtrsim 3$ in the
cosmological simulations IllustrisTNG}

\author[S. Kurinchi-Vendhan et al.]{Shalini Kurinchi-Vendhan,
$^{1,2}$
\thanks{E-mail: skurinch@caltech.edu
}
Marion Farcy,$^{3}$
Michaela Hirschmann,$^{3,4}$
Francesco Valentino$^{5,6}$
\\
$^{1}$ California Institute of Technology (Caltech), 1200 E.
California Boulevard, Pasadena, California 91125, United States of
America
\\
$^{2}$ Max-Planck-Institut für Astronomie (MPIA), Königstuhl 17,
Heidelberg 69117, Baden-Württemberg, Germany
\\
$^{3}$ École Polytechnique Fédérale de Lausanne (EPFL), Observatoire
de Sauverny, Chemin Pegasi 51, 1290 Versoix, Switzerland
\\
$^{4}$ Istituto Nazionale di Astrofisica (INAF), Osservatorio
Astronomico di Trieste, Via Tiepolo 11, 34131 Trieste, Italy
\\
$^{5}$ European Southern Observatory (ESO), Karl-Scwharzschild-Strasse
2, D-85748, Garching bei M\"unchen, Germany
\\
$^{6}$ Cosmic Dawn Center (DAWN), Niels Bohr Institute, Rådmandsgade
62-64, 2200 Copenhagen, Denmark
}

\date{Accepted 2024 September 30. Received 2024 September 26; in
original form 2023 October 2}
\pubyear{2024}

\begin{document}
\label{firstpage}
\pagerange{\pageref{firstpage}--\pageref{lastpage}}
\maketitle

\begin{abstract}
\noindent Using the cosmological simulations IllustrisTNG, we perform
a comprehensive analysis of quiescent, massive galaxies at $z \gtrsim
3$. The goal is to understand what suppresses their star formation so
early in cosmic time, and how other similar mass galaxies remain
highly star-forming. As a first-order result, the simulations are able
to produce massive, quiescent galaxies in this high-redshift regime.
We find that active galactic nuclei (AGN) feedback is the primary
cause of halting star formation in early, massive galaxies.  Not only
do the central, supermassive black holes (SMBHs) of the quenched
galaxies have earlier seed times, but they also grow faster than in
star-forming galaxies. As a result, the quenched galaxies are exposed
to AGN feedback for longer, and experience the kinetic, jet mode of
the AGN feedback earlier than the star-forming galaxies. The release
of kinetic energy reduces inflows of gas while likely maintaining
outflows, which keeps a low cold gas fraction and decreases the star
formation of the galaxies down to a state of quiescence. In addition
to AGN feedback, we also investigate the influence of the large-scale
environment. While mergers do not play a significant role in the
quenching process, the quenched galaxies tend to reside in more
massive halos and denser regions during their evolution. As this
provides a greater initial amount of infalling gas to the galaxies,
the large-scale environment can mildly affect the fate of the central
SMBH growth and, via AGN feedback, contribute to star formation
quenching.
\end{abstract}



\section{Introduction}  \label{sec:introduction} 

Exploring how star-forming galaxies are converted into passive, or
\textit{quenched}, galaxies is fundamental to our understanding of
galaxy evolution. The majority of massive galaxies become ``red and
dead'' over cosmic time, but the continuous discovery of massive
quenched galaxies at increasingly high-redshift remains puzzling
\citep[e.g.][]{Long_2023}. Some physical mechanisms may prevent gas
from being accreted onto the galaxy, keep gas in the galaxies from
cooling and condensing, or expel gas from the galaxies. Together,
these processes can deplete galaxies of their star-forming gas and
eventually shut down star formation. We provide a brief overview of
different possible quenching mechanisms in this introduction
\citep[see e.g.][]{Man_2018}. 

To sustain star formation, the gas reservoir of a galaxy has to be
regularly replenished by accretion from cosmic filaments or through
merger interactions with gas-rich companions. When a galaxy accretes
gas at an insufficient rate, this may eventually lead to cosmological
starvation \citep{Wetzel_2013}: the galaxy runs out of fuel and cannot
form stars. Furthermore, the accreted gas needs to reach the
interstellar medium (ISM) of the galaxy, which might be prevented if
gas cannot cool onto it. This can initially occur when virial shocks
contribute to heating the gas. Gravitational interactions with
satellites, which lead to dynamical friction, may also provide the
energy needed for increasing the temperature of the gas
\citep{Rees_Ostriker_1997, Khochfar_2008}. 

Once the gas reaches the ISM, a history of ``bursty'' star formation
and major mergers can facilitate the rapid loss of cold gas
\citep{Mihos_Hernquist_1996}. To a lesser extent, and in addition to
rapid star formation, active galactic nuclei (AGN) can consume cold
gas through accretion onto supermassive black holes (SMBHs),
eventually depleting the galaxy of part of its star-forming fuel. The
suppression of star formation activity is not only connected to the
gas reservoir but also depends on its ability to be converted into
stars. In other words, galaxies can quench when their gas does not
cool. While virial shocks and interactions with neighbouring galaxies
can temporarily stop gas from falling towards the galaxies, additional
mechanisms are required to prevent gas that reached the interstellar
medium (ISM) from cooling and condensing to form stars. Stellar
feedback from radiation, winds, and supernovae can lead to temporary
shock-heating \citep{Ciotti_1991,Springel&Hernquist_2003}, but is also
very likely to be inefficient in maintaining quiescence in massive
galaxies \citep[e.g.][]{Dubois_2016,Choi_2018}. Instead, it is largely
thought that sustained heating is driven by radiatively inefficient
accretion onto AGN \citep{Bower_2006, Croton_2006}: SMBHs accreting at
a low rate produce jets that inflate buoyant bubbles of gas, heating
the hot halo gas \citep{Fabian_2012}. \footnote{In some observations,
jets can also occur in radiatively efficient AGN as exceptions
\citep{Hardcastle_2018, Harcastle_Croton_2020}.} Other forms of
feedback can also play an important role in clearing out the gas and
quenching galaxies in the first place.

If gas is not prevented from cooling or condensing, a galaxy must
expel part of its reservoir of cold gas faster than it can be
replenished in order to remain quenched. Through near encounters with
surrounding galaxies in a dense environment, low-mass galaxies can
lose their star-forming gas content through processes such as
ram-pressure stripping, strangulation, and harassment
\citep{Hirschmann_2014b,Peng_2015, Book_2010, Boselli_2016,
Smethurst_2017}. As such, environmentally-driven processes might play
a direct role in affecting the galaxies’ cold gas reservoir. Stellar
and AGN feedback can also contribute to ejecting gas out of the
galaxies. On the one hand, when the most massive stars explode as
supernovae, the energy released can initiate galactic fountains
\citep{Chevalier&Clegg1985}. On the other hand, during episodes of
high accretion, the SMBH has enough energy to eject cold gas from the
galaxy through momentum-driven winds \citep{DiMatteo_2005,
Ishibashi_2012}. These winds can result from the radiation pressure of
the AGN as well as energy shocks from the accretion disk
\citep{Faucher-Giguere_2012, Thompson_2015, Ishibashi_2015,
Costa_2018}. As such, stellar and AGN feedback-driven winds can expel
enough gas from the galaxy, which limits star formation.

In this way, the causes of quenching can range from external to
internal processes which influence the gas reservoir in a galaxy and
thus suppress star formation. Whether cold gas does not enter the
galaxy, or the gas is heated, consumed, and removed at a faster rate
than it can cool, stellar feedback plays a subordinate role to
quenching. While environmental effects are thought to be the main
quenching channel for low mass galaxies ($M_\mathrm{stellar}$ below
$\simeq 10^{10}\,\mathrm{M}_\odot$, \citealp{Peng_2015}), AGN are
likely key to the long-term quiescence of massive systems. For
example, in low redshift observed galaxies, \cite{Terrazas_2017} show
that star formation rate is a decreasing function of SMBH mass,
establishing AGN as important to regulating the star formation in the
local Universe. At low redshifts, high-energy outflows from AGN can
contribute to star formation quenching through expelling gas from the
central regions of massive galaxies, as detected by ionized winds and
outflows of molecular gas in galaxies with AGN \citep{Feruglio_2010,
Fischer_2010, Sturm_2011, Westmoquette_2012, Rupke_2013, 
Harrison_2014}. There is also observational evidence for AGN feedback
at high redshifts, where nebular emission is found in broadline quasars
\citep{Arav_2013, Cano-Diaz_2012, Cicone2014, Genzel2014, 
ForsterSchreiber2014, Harrison_2015}. In principle, the energy output
from quasars should be able to deplete the gas reservoir of galaxies 
\citep{King_Pounds_2015}. However, it can be difficult to draw a causal
connection between AGN feedback and quiescence, since AGN feedback is 
so rare and highly variable, particularly in high-redshift 
observations. Moreover, AGN activity in such observations is often
coupled with recent mergers, making it challenging to isolate the exact
impact of SMBH feedback \citep{Schreiber_2019}. 

For determining whether AGN feedback is indeed responsible for
suppressing star-formation in galaxies, theoretical models and
simulations are a valuable informative tool. Semi-analytical models
(combined with $N$-body simulations or extended Press-Schechter based
merger trees) initially suggested the ability of AGN feedback to
create and maintain quenched massive galaxies, given the amount of
energy released by growing SMBHs
\citep{Croton_2006,Bower_2006,Somerville_2008}. Using subgrid models
for AGN feedback, numerical simulations also confirm that this process
is not only able, but is also required, to regulate star formation in
massive halos and reproduce the observed co-evolution between SMBH and
galaxy masses
\citep{DiMatteo_2005,Springel_2005,Sijacki_2007,Dubois_2012,Choi_2015}.
AGN feedback is hence one of the key ingredients to reproducing the
massive end of the galaxy stellar mass function, and is needed to
emulate quenched fractions of massive galaxies that agree with
low-redshift observations in cosmological simulations. These include
Magneticum \citep{Hirschmann_2014,Bocquet_2016}, Illustris
\citep{Vogelsberger_2014}, EAGLE \citep{Schaye_2015}, Horizon-AGN
\citep{Dubois_2016}, IllustrisTNG
\citep{Springel_2018,Pillepich_2018,Naiman_2018,Nelson_2018,Marinacci_2018,Nelson_2019}
and Simba \citep{Dave_2019}.

From the perspective of observations at $z = 3 - 4$,
\citet{Schreiber_2018, Merlin_2019, Girelli_2019, Cecchi_2019,
Valentino_2020, Santini_2021} all demonstrate the overall, qualitative
agreement of quiescent\footnote{In this work, we do not make any
distinction between quiescent and quenched galaxies, and we use both
words interchangeably.} massive galaxies with theoretical predictions
from most current hydrodynamic simulations, however with an increasing
tension at higher redshifts. Recently, a growing number of massive
quenched galaxies have been observed at $z\geq3$
\citep{Forrest_2020,Gould_2023, Antwi-Danso_2023}, especially with the
James Webb Space Telescope
\citep[JWST,][]{Nanayakkara_2022,Carnall_2023,
Carnall_2023Nature,Long_2023, Valentino_2023}, further confirming the
difficulty for cosmological simulations to recover the expected number
density of massive quenched galaxies at increasing redshifts. This
poses a challenge to our current understanding of star formation
activity in galaxy evolution, which struggles to capture the rapid
growth and sudden halt of the stellar mass assembly of galaxies as
massive as $M_\mathrm{stellar} = 10^{10} - 10^{11} \:
\mathrm{M}_\odot$ less than $\sim 2 \: \mathrm{Gyr}$ after the Big
Bang.

Because of previously limited and scarce observational data, there are
only a few theoretical works which study the causes of quenching in
massive galaxies beyond the local Universe
\citep{Hartley_2023,Lovell_2023,Lagos_2023}. Studies that focus on
$z\leq 2$ show that simulated quiescent and massive galaxies are
strongly influenced by AGN-driven energy injections
\citep{Donnari_2021a, Park_2022,Park_2023}. While there are hints that
AGN feedback already contributes to halting star formation in the
early Universe (e.g. \citealp{Hartley_2023} who focus on five massive
quenched galaxies in the IllustrisTNG simulation, or
\citealp{Lovell_2023} in the FLARES simulations at $z\leq5$), it
remains to be determined if galaxy mergers, which occur frequently at
high-redshift, and the large-scale environment, by affecting the
fueling of gas from filaments onto a halo, play any role in quenching
massive galaxies at early cosmic times. Therefore, disentangling
whether AGN feedback, merger events, and large-scale structure are
responsible for the quenching of massive galaxies at $z > 3$ is the
purpose of this study.

In this paper, we investigate the processes that lead to the
suppression of star formation in high-redshift, massive galaxies using
the cosmological simulation suite IllustrisTNG. Our method for
selecting and analyzing galaxies of interest is described in
\S~\ref{sec:methodology}. Then, \S~\ref{sec:results} looks at the
effect of AGN feedback, merger events, and large-scale structure in
causing quiescence. Next, \S~\ref{sec:discussion} discusses the
importance of the different quenching mechanisms in this study and in
other cosmological simulations, and states the limitations which may
explain why simulations are in tension with the observed number
densities of massive quiescent galaxies at $z\geq3$. We conclude in
\S~\ref{sec:conclusion} that AGN feedback is the most likely origin of
quiescent massive galaxies at high redshifts in IllustrisTNG, with
environmental conditions as an underlying contributing factor.

\section{Methodology}                \label{sec:methodology}

In this section, we describe the simulations, feedback models, and
selection procedure of the paper.

\begin{figure*}
\centering
\includegraphics[width=\linewidth]{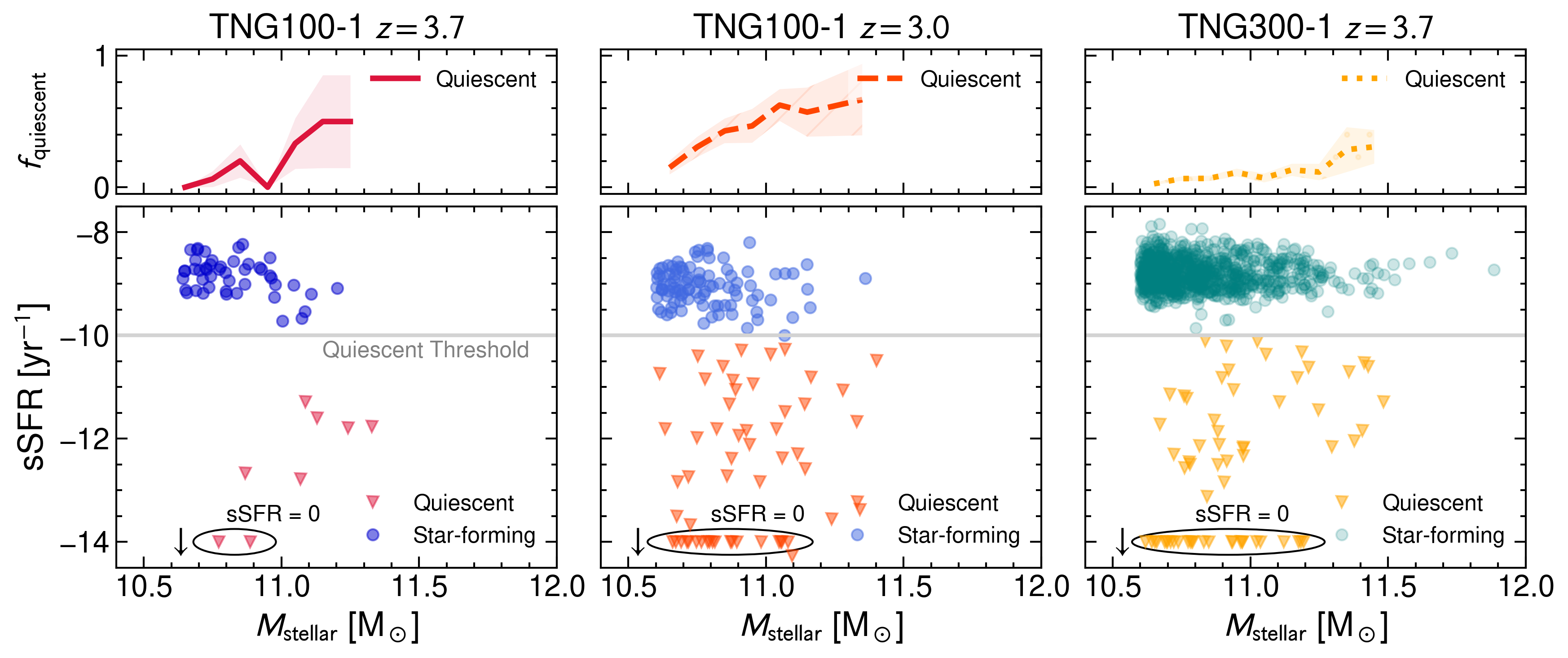}
\caption{\textbf{\textit{Quiescent and Star-forming Samples.}}
Fraction of quiescent galaxies (top row) and specific star formation
rate (bottom row) versus stellar mass for the TNG100-1 $z = 3.7$
(left), TNG100-1 $z = 3.0$ (middle), and TNG300-1 $z = 3.7$ (right)
samples. In each panel, we compare the quiescent (red shade,
triangles) and star-forming (blue shade, circles) galaxies. We select
galaxies above a mass-cut $M_\mathrm{stellar} \geq 10^{10.6} \:
\mathrm{M}_\odot$. Then, we define quiescent galaxies as having
specific star formation rates in the range $\mathrm{sSFR} \leq
10^{-10} \: \mathrm{yr}^{-1}$. The star-forming galaxies in our
selection have sSFR values above this threshold (gray, solid line).
Some galaxies in the quiescent samples have $\mathrm{sSFR} = 0 \:
\mathrm{yr}^{-1}$, as they have zero star-forming gas cells; these
objects are circled in black and indicated with a downward arrow. In
each simulation sample, there are more star-forming than quiescent
galaxies, and the fraction of quiescent galaxies increases with
stellar mass.}
\label{fig:sSFR_mass}
\end{figure*}

\subsection{IllustrisTNG simulations}\label{subsec:illustrisTNG}

We investigate the cause of quenching in high-redshift, massive
galaxies using the cosmological simulation suite, IllustrisTNG
\citep[TNG;][]{Nelson_2019}. These simulations are performed with the
{\sc{Arepo}} code \citep{Springel_2010}, which solves the equations of
magnetohydrodynamics and self-gravity using a moving-mesh technique.
The TNG simulations include various physical processes such as gas
cooling, star formation, stellar feedback and AGN feedback, as
described in detail by \citet{Weinberger_2016} and
\citet{Pillepich_2017}. 

We utilize the highest resolution realizations of the TNG100 and
TNG300 simulations \citep[TNG100-1 and TNG300-1 from][]{Springel_2018,
Pillepich_2018, Naiman_2018, Nelson_2018, Marinacci_2018}, which
consist of large cosmological boxes of volume 110$^3$ cMpc$^3$ and
302$^3$ cMpc$^3$, respectively. More specifically, TNG100 has $1820^3$
initial dark matter particles with a mass resolution of $\rm 7.5
\times 10^6 \, M_\odot$, and has a baryonic mass resolution of $\rm
1.6\times 10^6\,M_\odot$. The mass resolution in TNG300 is lower, and
reaches $\rm 5.9 \times 10^7 \, M_\odot$ for dark matter particles and
$\rm 1.1 \times 10^7 \, M_\odot$ for baryons. We use both simulations
because, while TNG100 provides better resolution, TNG300 allows for
the formation of more numerous and more massive galaxies due to its
larger volume.

\subsection{Star formation and stellar feedback} \label{subsec:star_formation}

The way in which galaxies become quiescent depends on \textit{how}
stars are able to form in the simulation. Due to resolution limits,
TNG treats star formation in a subgrid framework and relies on the
pressurization of a multi-phase ISM model. Stars stochastically form
from gas cells that reach above the density threshold $n_{\rm H} = 0.1
\: \rm{cm}^{-1}$, with pressurization from unresolved supernovae
included in the star-forming gas. As a consequence of this star
formation model, galaxies in the TNG simulations quench when there is
not enough dense gas. Taking into account additional criteria such as
gas temperature, velocity dispersion, or turbulence, would change the
star formation history and the quiescent state of the galaxies, but
such a study is beyond the scope of this paper. Feedback from star
formation is assumed to drive galactic-scale outflows in the form of
winds. However, we do not go into detail on the implementation of
stellar feedback since it has been shown to be ineffective toward
quenching massive galaxies in the literature.

\subsection{SMBH activity and AGN feedback} \label{subsec:black_hole}

Because we are especially interested in the role of central SMBHs in
suppressing star formation in massive high-redshift galaxies, we
briefly summarize how black holes are seeded, how they grow, and how
AGN feedback acts in the TNG simulations. More details can be found in
the reference papers from \citet{Vogelsberger_2013} and
\citet{Weinberger_2016}.

The TNG model places a black hole with seed mass of $M_\mathrm{seed} =
8 \times 10^5 \: h^{-1} \: \mathrm{M}_\odot$ in a friends-of-friends
(FoF) halo when it exceeds $M_\mathrm{FoF} = 5 \times 10^{10} \:
h^{-1} \: \mathrm{M}_\odot$. Here, FoF halo refers to a
gravitationally bound dark matter halo as determined by the SUBFIND
algorithm \citep{Springel_2001}. Haloes are then tracked in time using
merger trees that are obtained using the SubLink algorithm from
\cite{Rodriguez_Gomez_2015}. This algorithm identifies unique
descendants and chooses the main progenitors of each galaxy with the
``most massive history'' behind it.

Once a black hole is seeded, it grows at a rate which follows the
Bondi-Hoyle-Lyttleton accretion rate $\dot{M}_\mathrm{Bondi}$, limited
to the Eddington rate $\dot{M}_\mathrm{Edd}$. Additionally, a black
hole may grow by merging with another one during a galaxy merger
event. As a consequence from black hole accretion, energy is released
to the surrounding gas cells, which we refer to as AGN feedback.
Whether the energy is released as thermal or kinetic energy depends on
the Eddington ratio $f_\mathrm{Edd} = \dot{M}_\mathrm{Bondi} /
\dot{M}_\mathrm{Edd}$. This value describes how strongly a black hole
is accreting relative to its maximum possible accretion rate and is
dependent upon the mass of the black hole squared. In TNG, the
Eddington ratio is compared to a threshold called ``$\chi$'': whenever
the Eddington ratio is above this threshold and the accretion rate is
high, AGN feedback operates via the deposit of thermal energy
$E_\mathrm{inj, \: therm}\:$ in the quasar mode. Conversely, when
$f_\mathrm{Edd} < \chi$, the energy $E_\mathrm{inj, \: kin}\:$ is
imparted as random kinetic kicks in the so-called kinetic, radiatively
inefficient jet mode feedback. The original TNG methods paper for
black hole feedback \cite{Weinberger_2016} gives detailed definitions
of the above quantities, but we also provide them here for
completeness:
\begin{align}
\label{eq:1}
\dot{M}_\mathrm{Bondi} &= \frac{4 \pi G^2 M_\mathrm{BH}^2
\rho}{c_s^3}, \\
\label{eq:2}
\dot{M}_\mathrm{Edd} &= \frac{4 \pi G^2 M_\mathrm{BH} m_p
\rho}{\epsilon_r \sigma_T c}, \\
\label{eq:3}
\chi &= \min\left[\chi_0\times\left(\frac{M_{\rm BH}}{10^8\, \rm
M_\odot}\right)^\beta,0.1\right].
\end{align}
In Equations \ref{eq:1} and \ref{eq:2}, $G$ is the universal
gravitational constant, $m_p$ the proton mass, $\epsilon_r$ the
radiative accretion efficiency, $\sigma_T$ the Thompson cross-section,
and $c$ the vaccuum speed of light. $M_\mathrm{BH}$ is the black hole
mass, and $\rho$ and $c_s$ are the density and sound speed of the gas
near the black hole, respectively. In Equation \ref{eq:3}, $\chi_0 =
0.002$ and $\beta = 2$ which are the fiducial values adopted in TNG.
The functional form of $\chi$ is such that even the most massive black
holes in the simulation can reach the quasar feedback mode, while it
would be difficult for low mass black holes to reach the jet feedback
mode. This heuristic formulation of $\chi$ is meant to support the
quenching of massive galaxies. In the raw TNG data outputs, the
quasar-mode and jet mode feedback of the AGN are moreover given as
cumulative energy injections over time. We estimate the time-averaged
energy injection rate, or power $\dot{E}_\mathrm{inj, \: therm \: or
\: kin}\:$, as the change in $E_\mathrm{inj, \: therm}$ or
$E_\mathrm{inj, \: kin}$, divided by the difference in cosmic time
between snapshots.

It should be highlighted that the AGN feedback mechanisms are not
themselves resolved, and in general, cannot be resolved in
cosmological simulations which rely on subgrid models (for example,
~\protect\citealp{Schaye_2015} in EAGLE and
~\protect\citealp{Henden_2018} in FABLE too). Rather, the TNG works to
capture the large-scale impact of the quasar and jets through their
effect on the gas cells around the black hole. For example, TNG
constructs the jet mode to efficiently regulate the growth of massive
galaxies. This feature of cosmological simulations leads to
parameterizations which are somewhat numerically rather than entirely
physically motivated, in order to reproduce realistic galaxy
populations in the the local Universe. As a preliminary note, other
simulations use different subgrid feedback models to obtain similar
low-redshift galaxy populations: following the examples above, EAGLE
makes use of a single efficient thermal mode, and FABLE has a stronger
quasar mode than that of TNG together with the jet mode used in the
Illustris simulations \citep{Genel_2014, Vogelsberger_2014,
Sijacki_2015}. The choice of modeling will inherently impact our
interpretation of what quenches massive high-$z$ galaxies in these
simulations, which we further examine in \S~\ref{sec:discussion} of
the Discussion.

\subsection{Sample selection}\label{subsec:sample}

\begin{table}
\centering
\begin{tabular}{c|c|c|c}
Simulation     &   Redshift    &   \# Quiescent    &   \# Star-forming
\\
\hline
1. TNG100-1    &   $z = 3.7$   &   8               &   47
\\
2. TNG100-1    &   $z = 3.0$   &   56              &   99
\\
3. TNG300-1    &   $z = 3.7$   &   65              &   884
\\
\end{tabular}
\caption{\textbf{\textit{Selected Galaxies in TNG.}} We use three
different samples in this work, and include their respective numbers
of quiescent and star-forming galaxies here.}
\label{tab:1}
\end{table}

In order to study the high-redshift Universe, we first consider
simulated galaxies at redshift $z = 3.7$ in TNG100.  We only consider
resolved subhalos that have at least 1000 stellar particles and 1000
dark matter particles. Since this sample only relies on a single
simulation snapshot, we also select galaxies at $z = 3.0$ in TNG100 to
assess whether or not our findings are consistent across different
times in the high-redshift Universe. As a further check, we also
consider the $z = 3.7$ snapshot in TNG300 to obtain larger number
statistics. 

To select massive, quiescent objects, we make use of the stellar mass
and star formation content of the galaxies. We impose a mass-cut of
$M_\mathrm{stellar} \geq 10^{10.6} \: \mathrm{M}_\odot$, as adopted by
\citet{Valentino_2023} to compare the number densities of quenched
galaxies at $z=3-4$ in a compilation of observational works to that
measured in simulations. We note that throughout the paper, we
consider stellar and gas properties as the sum of all gravitationally
bound star and gas particles within twice the stellar half-mass radius
($R =$ 2 $R_{\rm eff}$) of the galaxy. Then, we define quiescent
galaxies as having specific star formation rates, or the ratio between
star formation rate and stellar mass, in the range $\mathrm{sSFR} \leq
10^{-10} \: \mathrm{yr}^{-1}$ (following e.g. \citealp{Franx_2008} for
$z\simeq3.7$). The star-forming galaxies in our selection have sSFR
values above this threshold.

This selection of galaxies based on stellar mass and star formation
rates is shown in Figure \ref{fig:sSFR_mass}. From left to right, we
show our three samples of galaxies from TNG100 at $z=3.7$, at $z=3.0$,
and from TNG300 at $z=3.7$. Quenched galaxies are represented with red
symbols, while star-forming galaxies are shown by blue symbols. We
adopt the same coloring scheme for all plots throughout the paper. The
top panels of Figure \ref{fig:sSFR_mass} show that the fraction of
quiescent galaxies ($f_{\rm quiescent}$) increases with stellar mass.
In TNG100, at $z=3.7$, only a few galaxies are quenched. After some
time, at $z=3$, lower-mass galaxies have built up: there are more
galaxies with $M_\mathrm{stellar} \geq 10^{10.6} \: \mathrm{M}_\odot$,
and more of them are quenched at this redshift than at $z=3.7$. TNG300
provides the larger sample for both quenched and star-forming massive
galaxies. At the high-mass end, only a handful of galaxies are
quiescent so that $f_{\rm quiescent}$ is smaller for this sample than
for those from TNG100 (in the latter, there are only two galaxies,
which is not statistically significant). The quiescent fractions are
overall lower in TNG300 than in the TNG100 samples, probably due to
resolution limits, and the fact that the stellar mass of halos of a
given total mass is slightly lower in TNG300 compared to TNG100
\citep{Pillepich_2018, Vogelsberger_2018}. This implies that there is
less feedback at the internal and environmental scales in the larger
simulation box when considering stellar mass bins. To better
appreciate the distribution of quiescent and star-forming galaxies,
the bottom panel of Figure \ref{fig:sSFR_mass} shows the specific star
formation rate as a function of stellar mass. The grey horizontal line
represents our quiescent threshold of $\mathrm{sSFR} = 10^{-10} \:
\mathrm{yr}^{-1}$. Overall, the number of quiescent galaxies is lower
than their star-forming counterparts in our samples. To summarize, 
the three different simulation samples consist of massive galaxies 
from Table \ref{tab:1}. Throughout this study, we will present the
findings for each of these three samples.


\section{Results}  \label{sec:results}

We consider three possible causes of quenching at high redshifts in
massive galaxies: internal SMBH feedback, mergers with other galaxies,
and the indirect influence of the large-scale environment. Later in
the discussion, \S~\ref{sec:discussion}, we explore the impact of
these quenching mechanisms on the gas reservoirs of the galaxies.

\subsection{Influence of AGN feedback} \label{subsec:BH_feedback}

First, we want to determine if feedback from the central SMBH is an
important cause of quiescence at high redshifts.

\subsubsection{SMBH mass growth}

As we mentioned in \S~\ref{subsec:black_hole}, in TNG, a black hole is
seeded in a galaxy once its host halo mass exceeds a certain mass,
$M_\mathrm{FoF} = 5 \times 10^{10} h^{-1} \: \mathrm{M}_\odot$. The
time at which this occurs in a galaxy is called the ``BH seed time,''
and gives information as to when the central SMBH starts growing and
releasing energy into the ISM. 

In order to have a better understanding of SMBH growth in our
galaxies, which may vary depending on the gas supply available, we
introduce the ``critical time of SMBH mass growth.'' We arbitrarily
define this as the time at which a SMBH first exceeds $M_\mathrm{BH}
\geq 10^{6.5} \: \mathrm{M}_\odot$ in mass. This happens relatively
soon after the black hole seeds and roughly corresponds to when the
black hole has doubled its initial mass, which is $M_\mathrm{BH} = 8
\times 10^{5} h^{-1} \: \mathrm{M}_\odot \approx 1.2 \times 10^{6} \:
\mathrm{M}_\odot$. This definition provides a more stable criterion
for detecting the seeding of black holes in the simulation.

\begin{figure*}
\centering
\includegraphics[width=\linewidth]{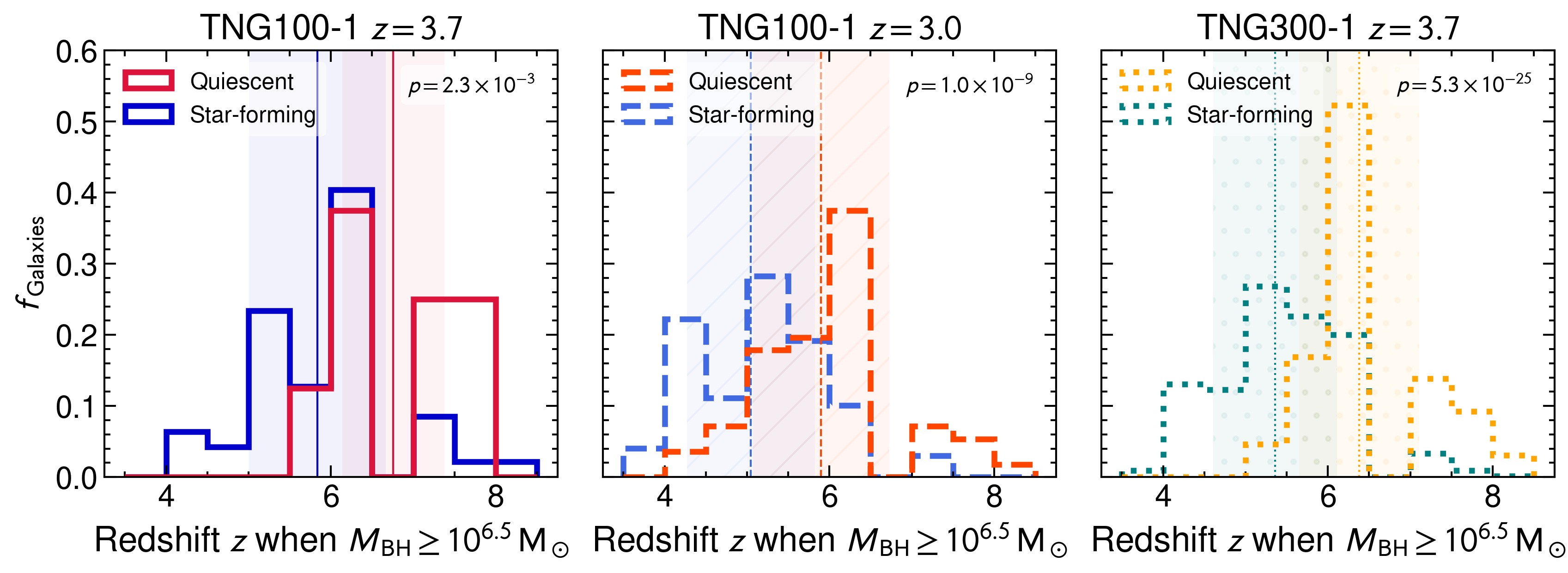}
\caption{\textit{\textbf{Impact of Black Hole Seeding.}} Relative
frequency of galaxies with respect to the critical time of SMBH mass
growth. We define the ``critical time of SMBH mass growth'' as the
time when the galaxy's central black hole mass exceeds $M_\mathrm{BH}
\geq 10^{6.5} \: \mathrm{M}_\odot$. We show separate histograms for
the TNG100-1 $z = 3.7$ (left, solid), TNG100-1 $z = 3.0$ (middle,
dashed), and TNG300-1 $z = 3.7$ (right, dotted) samples. We also
compare the histograms of the quiescent (red shades) and star-forming
(blue shades) galaxies in each panel, by indicating their
mean-averages (vertical lines) and standard-deviations (translucent
areas). In each simulation sample, the quiescent galaxies reach the
critical SMBH mass at significantly earlier times than the
star-forming galaxies (as also confirmed by the low $p$-values from
the one-sided T-test of the means, shown in the upper right
corners).}
\label{fig:bh_crit_hist}
\end{figure*}

Following the above definition, Figure \ref{fig:bh_crit_hist} shows
histograms of the redshift correponding to the critical time of SMBH
mass growth, for both the quiescent (red shades) and star-forming
(blue shades) samples across our three simulation sets. The thick
lines show the distribution of redshift, and the thin vertical lines
correspond to the average redshift, with the 1-$\sigma$ standard
deviation to the mean shown with a translucent area. The SMBHs of the
quiescent galaxies seed earlier and reach $M_\mathrm{BH} \geq
10^{6.5}\,M_\odot$ at earlier redshifts than the star-forming
galaxies. In the upper right corner of the plots, we show the
$p$-values from the one-sided T-test for the means of the two
independent galaxy samples. As they are low, this indicates that the
difference in the critical time of SMBH mass growth between the
star-forming and the quenched galaxies is statistically significant,
especially for the TNG300 sample due to its large number statistics.
Therefore, these earlier critical redshifts likely favor a faster SMBH
mass growth in the quiescent samples, and consequently a longer
exposure of the ISM to AGN feedback.

\subsubsection{Causes and consequences of SMBH growth and AGN
feedback}

\begin{figure*}
\centering
\includegraphics[width=\linewidth]{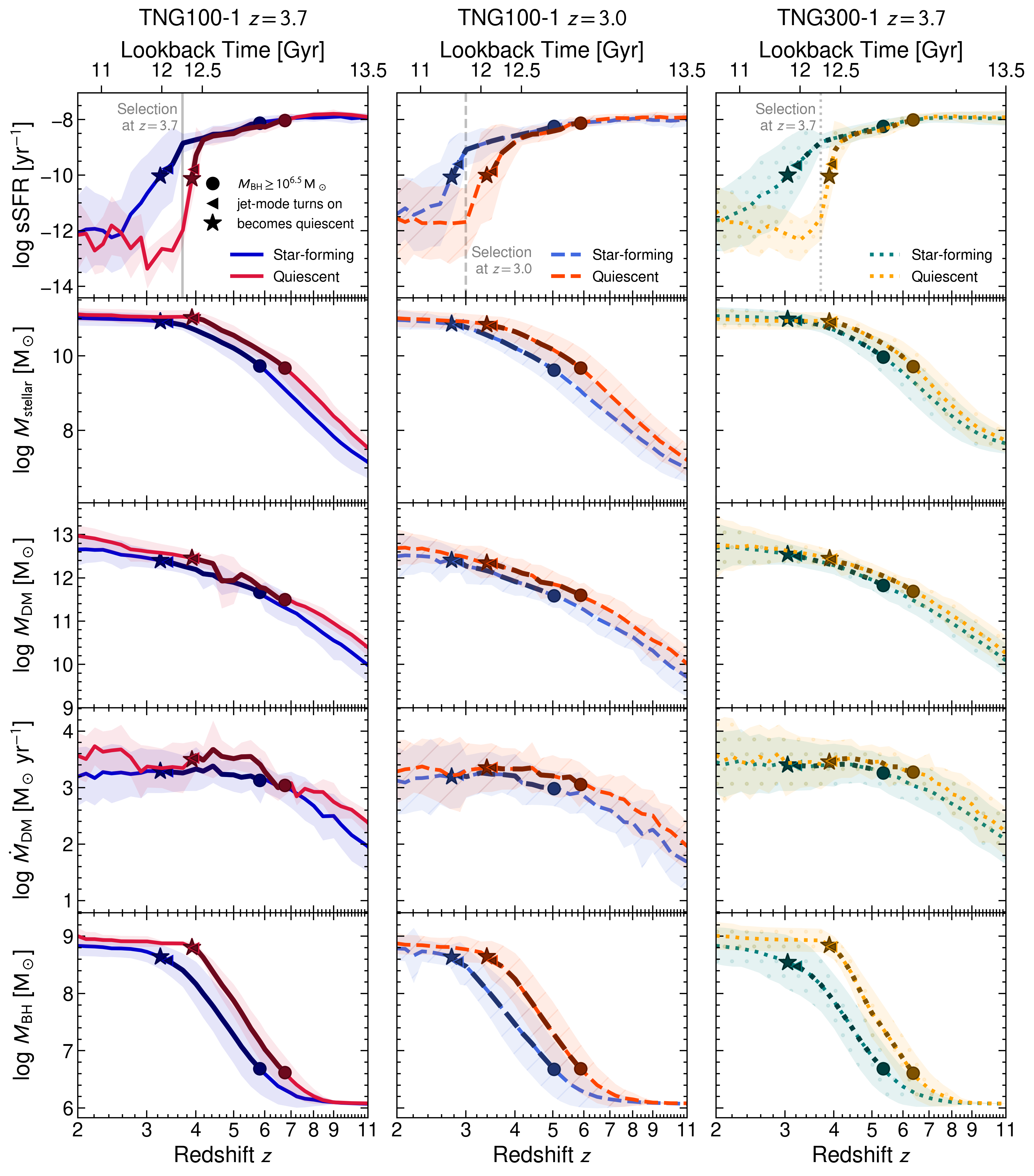}
\caption{\textit{\textbf{Stellar, Dark Matter, and SMBH Histories.}}
Redshift evolution between $z = 2$ to $11$ of various galactic
properties for the TNG100-1 $z = 3.7$ (left, solid), TNG100-1 $z =
3.0$ (middle, dashed), and TNG300-1 $z = 3.7$ (right, dotted) samples.
In each panel, we compare the quiescent (red shades) and star-forming
(blue shades) galaxies, by indicating their mean-averaged values and
standard-deviations (translucent regions). We also indicate the
average period (bolded segments) between when the galaxies' SMBH
masses first exceeded $10^{6.5} \: \mathrm{M}_\odot$ (circle), when
their AGN switched to the jet mode (triangle), and the time of
quiescence (star). From top to bottom, we show the logarithms of
specific star formation rate (sSFR), stellar mass ($M_{\rm stellar}$),
halo dark matter mass ($M_{\rm DM}$), dark matter accretion rate
($\dot{M}_{\rm DM}$), and SMBH mass ($M_{\rm BH}$). In general, the
sSFR (first row) of quiescent galaxies decreases below that of the
star-forming galaxies before $z = 4.0$. Before they begin quenching,
the quiescent galaxies also experience earlier and faster stellar mass
growth (second row). Because they are hosted in more massive halos
(third row), the quiescent galaxies experience higher dark matter mass
accretion rates (fourth row). Together, the stellar and dark matter
mass growth likely drive the accelerated mass growth of the galaxies'
central SMBHs (fifth row).}
\label{fig:basic_trees}
\end{figure*}
\begin{figure*}
\ContinuedFloat
\centering

\includegraphics[width=\linewidth]{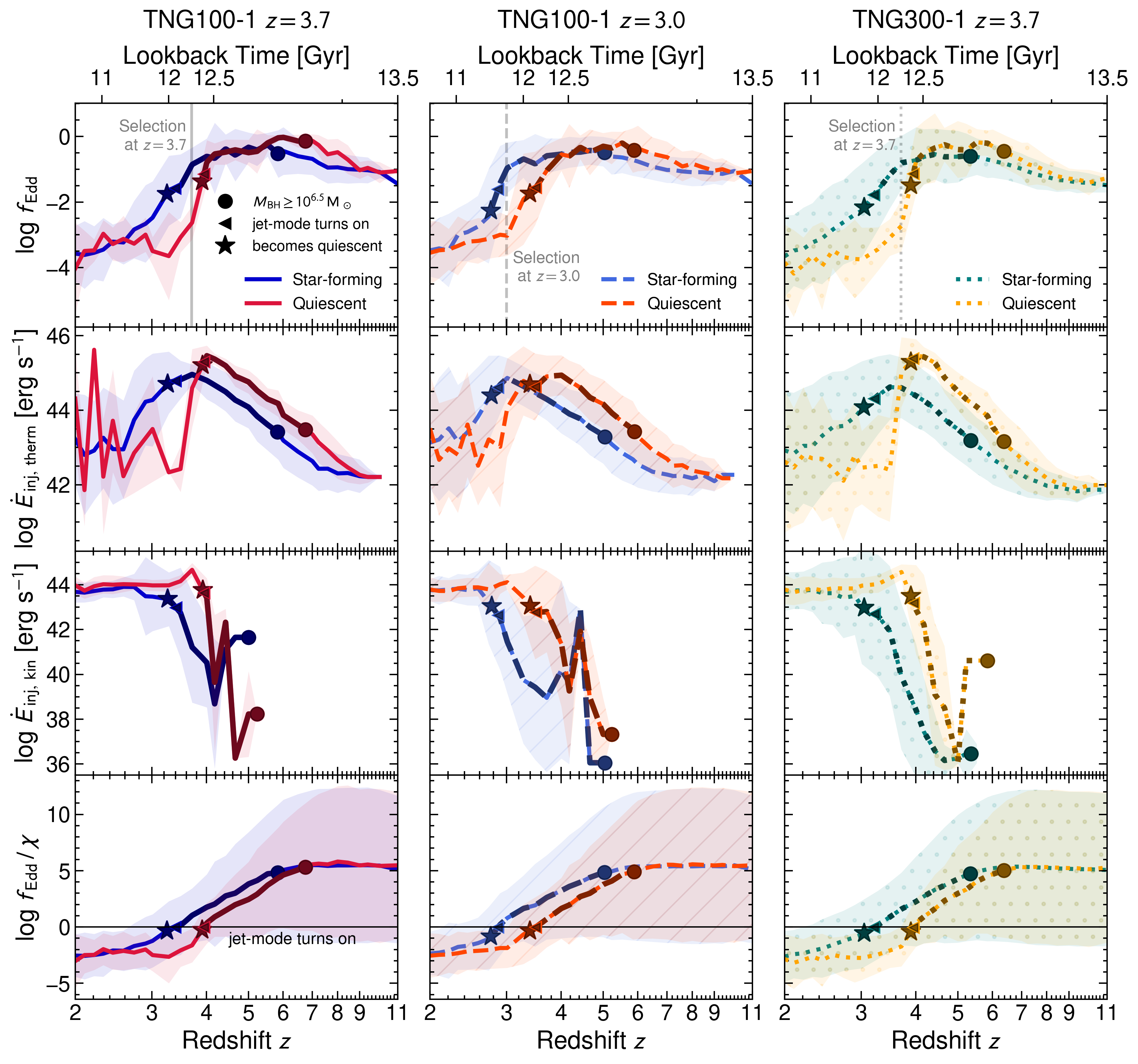}
\caption{ \textit{\textbf{(continued)}} Redshift evolutions of
Eddington ratio $f_\mathrm{Edd}$, AGN thermal energy injection rate
($\dot{E}_{\rm inj,\ therm}$), AGN kinetic energy injection rate
($\dot{E}_{\rm inj,\ kin}$), and ratio between the Eddington ratio and
the threshold for thermal to kinetic energy injection $\chi$. The
horizontal line in the bottom panels show the limit above (below)
which thermal (kinetic) energy is injected from the AGN.  While $\log
f_{\rm Edd}/\chi>0$, most of the energy injected by the SMBH is in the
thermal form (by definition). Before the time of quenching, SMBHs in
the quiescent galaxies have higher accretion rates, thus a higher
Eddington factor, and inject more energy in the thermal form. Then,
the kinetic jet mode of AGN feedback is triggered---which happens
slightly before, if not at, the time of quiescence---and from this
point, kinetic energy injections dominate the AGN feedback energy
budget. All of these results appear to be consistent between each
simulation sample.}
\end{figure*}

The consequences from SMBH growth and feedback taking effect earlier
in the quiescent galaxies are shown in detail in Figure
\ref{fig:basic_trees}. This figure shows the mean-averaged time
evolution of stellar, dark matter, and SMBH properties of the
simulated galaxies. From the top row to the bottom row, we
respectively focus on the specific star formation rate, stellar mass,
dark matter mass, dark matter accretion rate\footnote{We calculate
$\dot{M}_\mathrm{DM}$ using the net change of the dark matter mass
over the time-difference between simulation snapshots.}, SMBH mass,
Eddington ratio, AGN thermal energy injection rate, AGN kinetic energy
injection rate, and ratio between the Eddington ratio and $\chi$
(defined in \S~\ref{subsec:black_hole}). We note that dark matter
properties are calculated from all particles of this type within the
subhalo. We additionally indicate the time at which our samples of
galaxies are selected with grey vertical lines, as well as the period
between the critical time of SMBH mass growth, the onset of the AGN
jet mode, and the time of quiescence\footnote{The time of quiescence
for the star-forming samples corresponds to an average value which
does not include galaxies with $M_{\rm stellar}<10^{10.6}\,M_\odot$
that will reach this stellar mass later, after our selection time.}
with respective markers.

On average, the specific star formation rates of the quiescent and
star-forming galaxies are comparable until they diverge just before $z
= 4.0$, which is approximately when we select our samples. At this
point, the sSFRs of the quiescent samples experience a sharp drop
below that of the star-forming samples. The purpose of this subsection
is to investigate if this sudden decrease in star formation is related
to AGN feedback.

Continuing with Figure \ref{fig:basic_trees}, we see that the average
stellar mass of the quiescent galaxies is consistently higher than
that of the star-forming galaxies since the beginning of their
evolution. This would suggest higher SMBH masses according to known
scaling relations \citep{Kormendy_2013}, in line with the trend we
actually measure. We also highlight that quiescent galaxies are
overall hosted in more massive halos by $\sim 0.4 \: \mathrm{dex}$,
and consequently have higher dark matter mass accretion rates,
particularly for the TNG100 $z = 3.7$ and $z = 3.0$ samples. 

As a consequence from their stronger gravitational attraction (and
because they reside in high-density peaks, see also
\S~\ref{subsec:environment}), the quiescent galaxies also initially
have more gas supply, which favours star formation and SMBH growth.
The halos hosting the quiescent galaxies become massive enough to host
a SMBH earlier than their star-forming counterparts (when
$M_\mathrm{FoF} > 5 \times 10^{10} \: h^{-1} \: \mathrm{M}_\odot$), as
already shown with the critical time of SMBH mass growth in Figure
\ref{fig:bh_crit_hist}. Indeed, when we check narrower halo mass and
stellar mass bins, the time difference between when black holes seed
in the quiescent and star-forming samples becomes less significant.
Having their black hole sink particle seeded earlier is another
explanation for the faster SMBH growth for the quiescent galaxies,
which we further illustrate via the Eddington ratio: the SMBH
accretion rate experiences a higher peak in the quiescent galaxies
just before their SMBHs reach our defined critical mass of
$M_\mathrm{BH} \geq 10^{6.5} \: \mathrm{M}_\odot$, and until their
sSFRs begin to decrease. When the galaxies become quiescent, the
Eddington ratio decreases and the SMBH mass almost stops growing.

In order to better study the interplay between AGN feedback and star
formation, we now focus on the energy released by the SMBHs as a
consequence from their accretion. As discussed in
\S~\ref{subsec:BH_feedback}, in TNG, this energy is released either as
thermal or kinetic energy if the Eddington ratio is higher or lower
than a value $\chi$ (defined in \S~\ref{subsec:black_hole}),
respectively. 

From the last three bottom rows of Figure \ref{fig:basic_trees}, we
see that at early stages, from the time that the black hole seeds up
to the time of quiescence, AGN feedback mainly consists in thermal
energy injections. To remind the reader, we calculate the
time-averaged energy injected by dividing the change in energy by the
difference in cosmic time between snapshots. SMBHs hosted in the
quiescent galaxies are overall more massive than those hosted in
galaxies from our star-forming samples, and as a result they
contribute to more thermal energy injections. This is because the
total energy injected is by definition proportional to the SMBH
accretion rate, and hence scales with the SMBH mass squared. The
turnover time at which the AGN goes from a thermal, quasar mode to a
kinetic, jet mode is also a function of SMBH mass squared (via
$\chi$), and therefore SMBHs from the quiescent galaxy samples fall in
the kinetic feedback regime earlier than star-forming galaxies. 

Thus, on average, the quiescent galaxies in the TNG simulations
experience a greater injection of kinetic energy, which has been shown
to be the most efficient feedback channel to affect massive galaxies
and their star formation at low redshifts
\citep[e.g.][]{Terrazas_2020}. We also note that in our three
simulation sets, the time at which the kinetic jet mode of AGN
feedback is triggered coincides with the time of quiescence. From this
population-based analysis, it is not immediately clear whether the
galaxies become quenched first and the jet mode simply maintains their
quiescence, or vice versa. Zoom-in simulations of individual galaxies
are required, and initial explorations indicate that the AGN jets are
indeed driving the quenching process (Farcy et al. in preparation).
Nevertheless, it is apparent that SMBH activity is likely connected to
the drop in star formation of these massive galaxies.

From this first analysis, we conclude that in addition to having
earlier black hole seed times on average, the greater SMBH activity in
the quiescent galaxies must be driven by progenitor bias from higher
initial halo masses. At a given time, the SMBHs from our samples of
quenched galaxies are more massive, have a stronger accretion
activity, and inject more energy to the surrounding gas. Once the
SMBHs are massive enough and at the same time have an accretion rate
small enough, they trigger the kinetic, radiatively inefficient AGN
feedback mode, which coincides with a drop in sSFR and with the time
of quiescence.

\subsection{Negligible role of merger events} \label{subsec:mergers}

We just showed that SMBH feedback plays an important role in quenching
massive galaxies earlier in their lifetimes. Now, we focus on the
potential effect of merger events. 

The internal properties of a galaxy can be directly impacted by close
encounters and interactions with nearby subhalos. We investigate the
effect of minor and major merger events on internal galactic
properties through tracing the main branch of the galaxies' merger
trees back in time. This allows us to determine if quiescent and
star-forming galaxies experience different merger histories, and,
accordingly, if their stellar, gas, and SMBH evolutions may be
related. In our analysis, we define ``minor'' mergers as merger events
between galaxies with a mass ratio between 1:10 and 1:4. We identify
``major'' merger events as having a minimum mass ratio of 1:4. In
other words, we consider that a given galaxy has experienced a major
(minor) merger if it became the main progenitor of a galaxy that was
up to 4 times (between 4 and 10 times) less massive.

\begin{figure*}
\centering
\includegraphics[width=\linewidth]{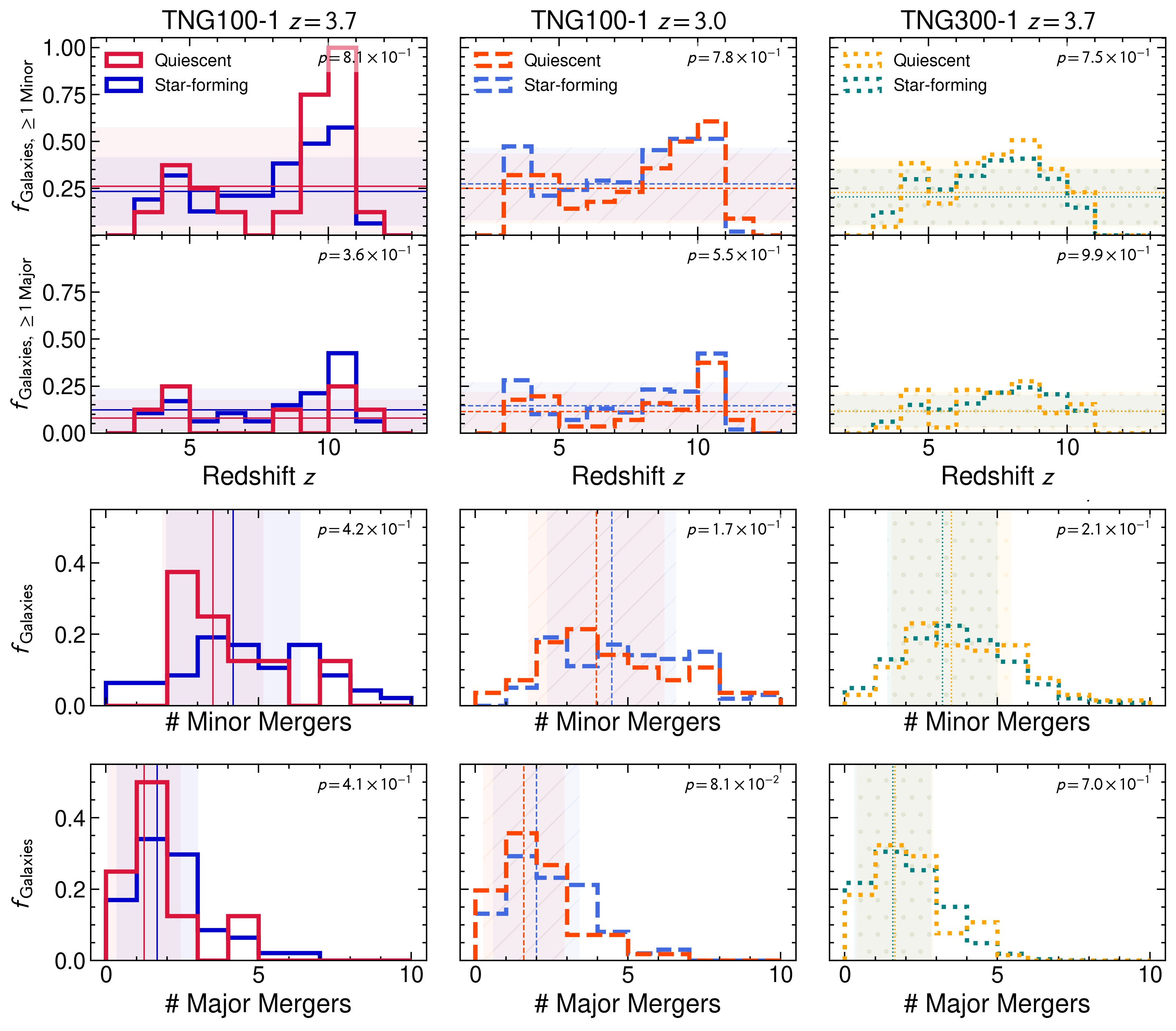}
\caption{\textit{\textbf{Impact of Minor and Major Mergers.}}
Histograms of the fraction of galaxies that experience merger  events
in the TNG100-1 $z = 3.7$ (left, solid), TNG100-1 $z = 3.0$ (middle,
dashed), and TNG300-1 $z=3.7$ (right, dotted) samples. We also compare
the histograms of the quiescent (red shades) and star-forming (blue
shades) galaxies in each panel, by indicating their mean-averages
(horizontal and vertical lines) and standard-deviations (translucent
boxes). \textit{First and second rows}: fraction of galaxies with
respect to redshift that experience at least one minor or major merger
with minimum mass ratios of $> 1:10$ and $> 1:4$, respectively. For
each simulation sample, the distribution of mergers over time is
similar between the quiescent and star-forming galaxies. \textit{Third
and fourth rows}: fraction of galaxies with respect to the number of
past minor and major merger events. The quiescent galaxies do not
experience significantly more mergers than the star-forming galaxies
(as also confirmed by the $p$-values approaching unity, shown in the
upper right corners). In this way, mergers between galaxies are likely
unimportant to the quenching of these galaxies at high-redshift.}
\label{fig:merger_hist}
\end{figure*}

Figure \ref{fig:merger_hist} shows the fraction of quiescent and
star-forming galaxies that experience each type of merger as a
function of redshift (first two rows), and as a function of the total
number of mergers at the redshift when galaxy samples are selected
(last two rows). 

Focusing first on the redshift evolution, we see that the overall
distribution of major and minor merger events is not significantly
different between the quiescent and star-forming galaxies. Although
the quiescent galaxies in the TNG100 $z = 3.7$  sample experience a
higher peak in the minor merger events at $z\simeq10$, this feature
does not appear in the TNG100 $z = 3.0$ and TNG300 $z = 3.7$
simulation sets. The mean fraction of both minor and major mergers,
shown with thin horizontal lines, is very similar for both populations
of galaxies, revealing that they have very similar merger histories.
In addition, the two bottom rows of Figure \ref{fig:merger_hist} show
that the quiescent galaxies do not experience significantly more minor
or major mergers than the star-forming galaxies. This is further
confirmed by the fact that the $p$-values approach unity, as shown in
the top right corner of the plots. If we look at the average number of
merger events, shown with thin vertical lines, most quiescent and
star-forming galaxies have experienced less than three minor mergers,
and only around one major merger, by the time we ``observe'' them in
the specified snapshots.  Overall, this shows that there is no excess
of recent major nor minor mergers in the quiescent galaxies.
Therefore, the IllustrisTNG simulations do not predict that the
quiescence of high-redshift massive galaxies is related to any
merger-driven starbursts that would have depleted the galaxies from
most of their star-forming, cold gas reservoirs \citep{Man_2018}.

\subsection{Large-scale environment}
\label{subsec:environment}

In \S~\ref{subsec:BH_feedback}, we explored the impact of SMBH growth
and AGN feedback on star formation quenching in high-redshift massive
galaxies, and in \S~\ref{subsec:mergers}, we also assessed whether
mergers are an external cause of quenching. In this section, we will
investigate the influence of the large-scale environment: that is, if
the location of galaxies within the cosmic web may favour their
quiescence. \\ \\ \\

\subsubsection{Residence time in a massive host halo}

One way to characterize the environment of the galaxies is through
looking at the dark matter content of their host halos, namely the
mass $M_\mathrm{host}$. In particular, one of the conclusions from
Figure \ref{fig:basic_trees} is that quenched galaxies are globally
hosted in halos more massive than star-forming galaxies. 

\begin{figure*}
\centering
\includegraphics[width=\linewidth]{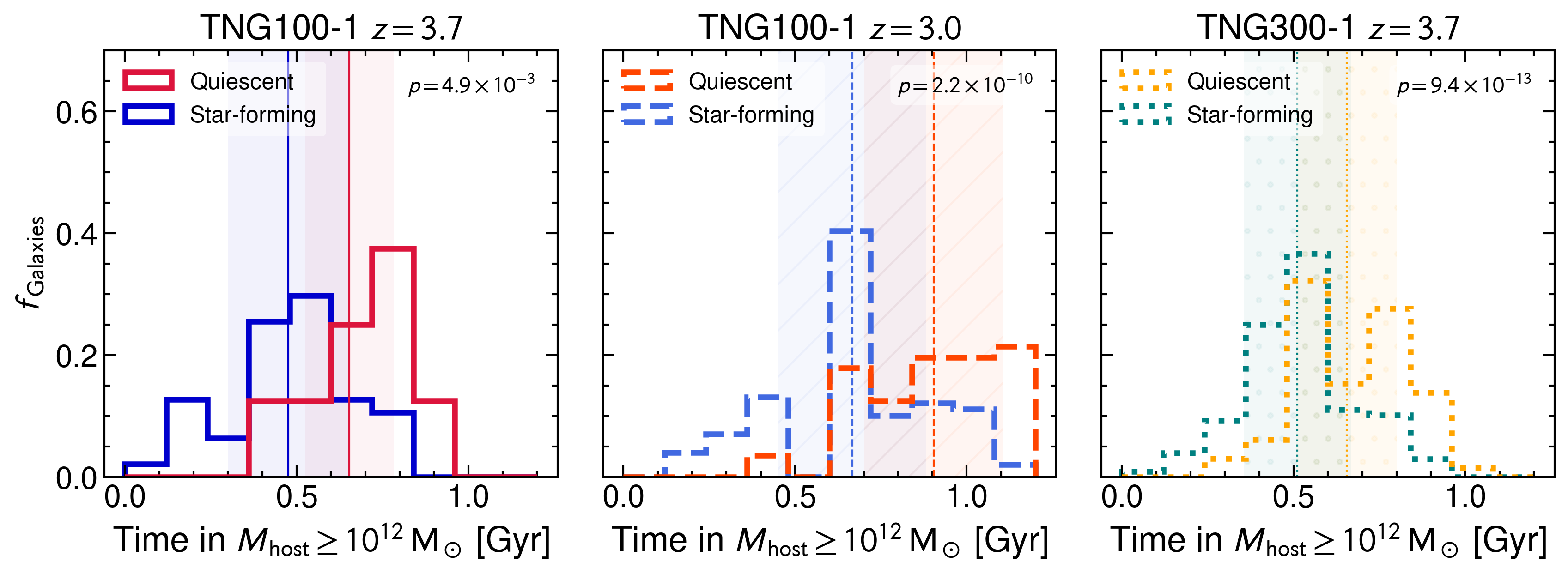}
\caption{\textit{\textbf{Impact of Massive Host Halos.}} Relative
frequency of galaxies with respect to the amount of time that they
reside in a ``massive host halo'', which we define as halos with a
dark matter mass $\geq 10^{12} \: \mathrm{M}_\odot$. We show separate
histograms for the TNG100-1 $z = 3.7$ (left, solid), TNG100-1 $z =
3.0$ (middle, dashed), and TNG300-1 $z = 3.7$ (right, dotted) samples.
We also compare the histograms of the quiescent (red shades) and
star-forming (blue shades) galaxies in each panel, by indicating their
mean-averages (vertical lines) and standard-deviations (translucent
box). In each simulation sample, the quiescent galaxies spend more
time in a denser environment (i.e. they are hosted in  massive halos
for longer) than the star-forming galaxies (as also confirmed by the
low $p$-values from the one-sided T-test of the means, shown in the
upper right corners).}
\label{fig:host_time_hist}
\end{figure*}

To complement this result, we determine if the quenched galaxies have
different large-scale environmental conditions through time compared
to star-forming galaxies by computing the amount of time they spend in
host halos more massive than $M_\mathrm{host} \geq 10^{12} \:
\mathrm{M}_\odot$. Whenever the subhalo is a satellite
galaxy,\footnote{Only a small percentage of our selected galaxies are
satellites, whereas the vast majority are centrals as expected from a
hierarchical structure formation scenario.} we will consider the dark
matter mass of its host halo. Figure \ref{fig:host_time_hist} shows
histograms of the relative frequency of quiescent and star-forming
galaxies with respect to the total amount of time they have been in
massive host halos, for all simulation sets. On average, the quiescent
galaxies spend statistically more time in a massive host halo than
star-forming galaxies, as indicated by the very small $p$-values.
However, the right tail of the blue histograms shows that there is a
sub-population of star-forming galaxies which spend a similarly large
amount of time in a massive host halo. Thus, a dense environment could
be a necessary, but not sufficient, condition for quiescence at
high-redshift.

\subsubsection{Number of neighboring satellite galaxies}

\begin{figure*}
\centering
\includegraphics[width=\linewidth]{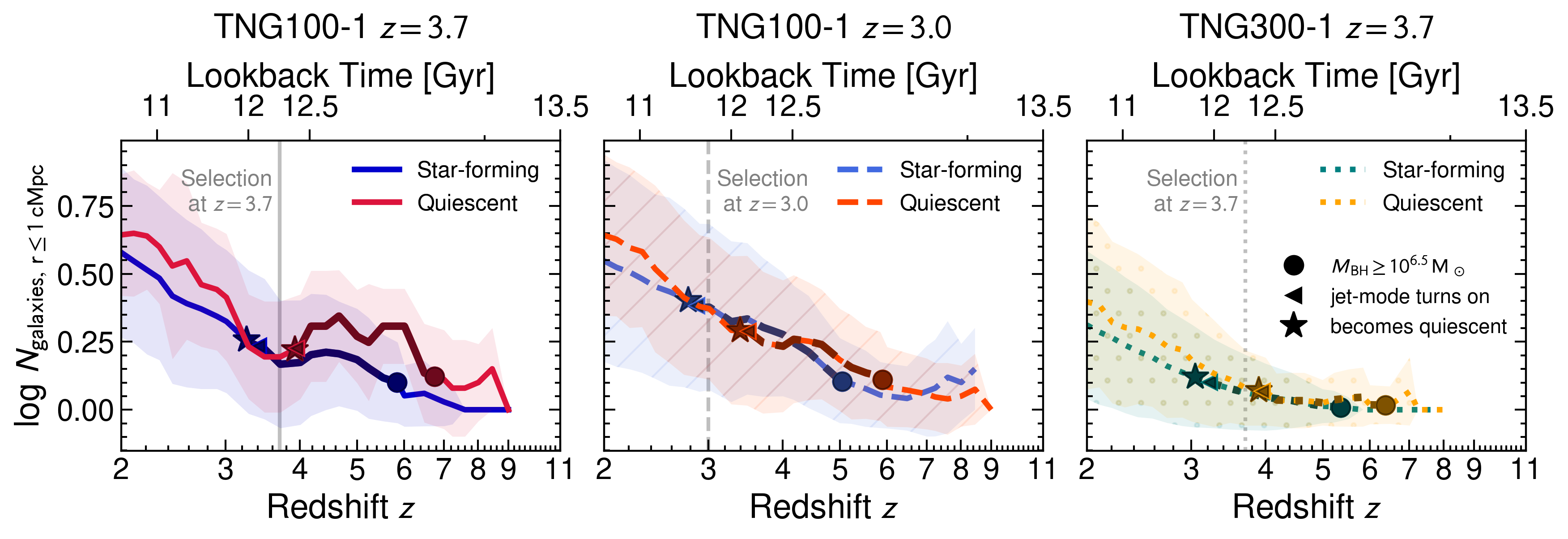}
\caption{\textit{\textbf{Environmental Histories.}} Redshift
evolutions between $z = 11$ to $2$ of the environmental density for
the TNG100-1 $z = 3.7$ (left, solid), TNG100-1 $z = 3.0$ (middle,
dashed), and TNG300-1 $z = 3.7$ (right, dotted) samples. In each
panel, we compare the quiescent (red shades) and star-forming (blue
shades) galaxies, by indicating their mean-averaged evolutions and
standard-deviations (translucent regions). We also indicate the
average period (bolded segments) between when the galaxies' SMBH
masses first exceeded $10^{6.5} \: \mathrm{M}_\odot$ (circle), when
their AGN switched to the jet mode (triangle), and the time of
quiescence (star). We calculate the environmental density of a galaxy
using the number of resolved neighboring subhalos within 1 cMpc.
Subhalos are considered to be resolved if they have at least a
thousand stellar particles and a thousand dark matter particles. In
general, the quiescent galaxies are in slightly more overdense
regions. In this way, dense environmental conditions may impact galaxy
quenching at high-redshift. The above results appear to be mostly
consistent between each simulation sample, except that the TNG300-1
galaxies tend to have lower average densities due to more stringent
dark matter and stellar mass cuts.}
\label{fig:density_trees}
\end{figure*}

The host halo mass provides a general sense of the environment that a
galaxy is in. As an additional check, we also measure the
environmental ``over''-density through counting the number of
neighboring satellites within 1 cMpc of each galaxy of interest. As
stated in \S~\ref{subsec:sample}, we only consider resolved subhalos
that have at least 1000 stellar particles ($M_\mathrm{stellar}$ $\geq
1.4\times10^{9} \: \mathrm{M}_\odot$ for TNG100 and $\geq
1.1\times10^{10} \: \mathrm{M}_\odot$ for TNG300) and 1000 dark matter
particles ($M_\mathrm{DM}$ $\geq 7.5 \times 10^{9} \:
\mathrm{M}_\odot$ in TNG100 and $\geq 5.9\times10^{10} \:
\mathrm{M}_\odot$ in TNG300). Figure \ref{fig:density_trees} shows the
time evolution of the average number of neighbours within a 1 cMpc
sphere centered around the quiescent and star-forming objects, in the
same style as Figure \ref{fig:basic_trees}. Overall, the quiescent
galaxies reside in slightly more dense regions, especially before and
after the SMBH mass growth-to-quenching period. Because we only count
the number of galaxies well resolved in terms of dark matter and
stellar content, we identify only a few neighbours around each target
halo, and their number increases with decreasing redshift as more
subhalos reach the dark matter and stellar mass cuts. 

Together with Figure \ref{fig:host_time_hist}, Figure
\ref{fig:density_trees} shows that quiescent massive galaxies sit in
higher density peaks compared to star-forming galaxies. Therefore,
quenched galaxies from our samples are able to evolve faster as they
have more gas accretion along filaments, leading to faster galaxy and
SMBH growth (as shown in Figure \ref{fig:basic_trees}). Because SMBH
growth is related to AGN feedback, this may be responsible for
shutting down star formation in the end. 

\section{Discussion}  \label{sec:discussion}

In the previous section, we put emphasis on three potential causes for
quenching massive galaxies at $z\gtrsim3$: SMBH growth and activity,
excess of minor and major mergers, and large-scale environmental
conditions. We now want to explore their consequences on the galaxies'
gas content and flows in \S~\ref{subsec:gas_flows}, compare our
results with those from other studies and at different redshifts in
\S~\ref{subsec:comparison}, and discuss the limitations of our work in
\S~\ref{subsec:caveat}.

\subsection{Consequences of feedback and environment} \label{subsec:gas_flows}

\begin{figure*}
\centering
\includegraphics[width=\linewidth]{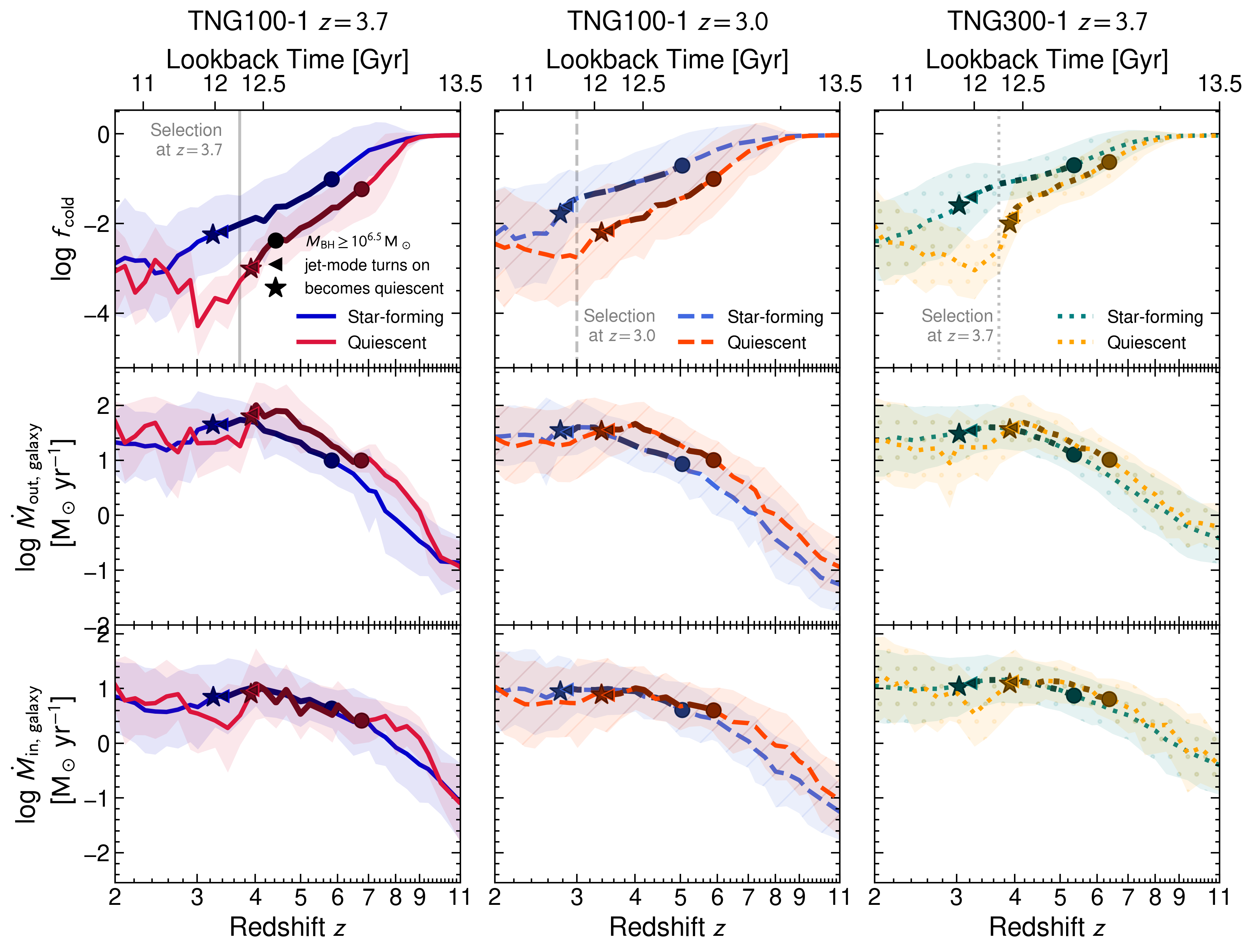}
\caption{\textit{\textbf{Gaseous Histories.}} Redshift evolutions
between $z = 11$ to $2$ of cold gas fraction (top row), gas outflow
rate (middle row) and gas inflow rate (bottom row) for the TNG100-1 $z
= 3.7$ (left, solid), TNG100-1 $z = 3.0$ (middle, dashed), and
TNG300-1 $z = 3.7$ (right, dotted) samples. In each panel, we compare
the quiescent (red shades) and star-forming (blue shades) galaxies, by
indicating their mean-averaged evolutions and standard-deviations
(translucent regions). We also indicate the average period (bolded
segments) between when the galaxies' SMBH masses first exceeded
$10^{6.5} \: \mathrm{M}_\odot$ (circle), when their AGN switched to
the jet mode (triangle), and the time of quiescence (star). Gas
fractions are measured at twice the stellar half-mass radius, and mass
flow rates are measured in spherical shells located at $R_{\rm galaxy}
= 7 R_\mathrm{eff}$ (far away enough from the galaxies while still in
the circumgalactic medium). Starting at $z \sim 8$, the cold gas
fraction of the quiescent galaxies begins to decrease below that of
the star-forming galaxies. Until their quenching time, quenched
galaxies have inflow and outflow rates higher than star-forming
galaxies. At the quenching time, quenched galaxies have a small drop
in inflow rate and still produce outflows despite a very weak stellar
feedback.}
\label{fig:gas_trees}
\end{figure*}

The gas content within and surrounding a galaxy can provide
information about how its star-forming reservoir is influenced by
internal processes like stellar and AGN feedback, versus the
environment. We now want to assess the effect of the large-scale
environment and of feedback on the galaxies' gas content and flows
through looking at their cold gas fraction, outflows, and inflows. 

In order to characterize the amount of cold gas in a galaxy, we
calculate the cold gas fraction as $f_\mathrm{cold} = M_{\mathrm{cold
\: gas}} / (M_\mathrm{cold \: gas} + M_\mathrm{stellar})$ following
definitions by observers, where $M_{\mathrm{cold \: gas}}$ is the mass
of all gas particles which have a non-zero star formation rate and a
temperature $T \leq 10^{5} \: \mathrm{K}$, and that belong to a sphere
of twice the stellar half-mass radius ($R = 2 R_\mathrm{eff}$). Only a
small fraction of star-forming gas cells on the effective equation of
state are above this temperature \citep[for reference,
see][]{Torrey_2019}.

In addition, we consider the outflow and inflow rates of gas through
the galaxy. To calculate the mass flow rates, we draw spherical shells
centered on each galaxy, of radius $R_{\rm galaxy} = 7 R_\mathrm{eff}$
and of $\rm 1\,ckpc$ width.\footnote{We estimate the radius of the
galaxy based on \cite{Kravtsov_2013} and \cite{Somerville_2018}, where
the effective radius of the galaxy is found to be $R_{\rm eff} = 0.015
R_{\rm halo}$ in terms of the halo virial radius for massive galaxies.
Assuming the entire extent of the galaxy is a tenth of the halo
\citep{Oser_2010}, we obtain $R_{\rm galaxy} \equiv R_{\rm halo} / 10
\simeq 7 R_{\rm eff}$. We also computed mass flow rates within a shell
of 10 kpc in width, and with a radius of $R_{200}$, and the results
are barely impacted.} For each gas cell that belongs to this shell,
the mass flow rates are then defined as the product of the gas mass
with the gas radial velocity divided by the width of shell. Gas cells
are considered inflowing (respectively outflowing) when they have a
positive (negative) radial velocity. The total mass inflow and outflow
rates are then derived by summing the values of all the cells
intersected by the shells.

Figure \ref{fig:gas_trees} shows the time evolution of the cold gas
fraction, the mass outflow rate, and the mass inflow rate (from top to
bottom). Before $z\simeq8$, the cold gas fraction does not largely
differ between the quiescent and star-forming galaxies, and before
their SMBH seeds, quiescent galaxies experience greater gas mass
inflow rates than the star-forming galaxies. This can be explained by
the fact that quiescent galaxies are hosted in more massive halos
(Figure \ref{fig:basic_trees}) and are surrounded by a slightly denser
environment (Figure \ref{fig:density_trees}), and therefore have more
gas supply available. With time, this enables stars to form and SMBHs
to grow at a faster rate in galaxies from the quiescent samples.
Because of higher star formation and SMBH accretion rates, supernova
feedback episodes and continuous SMBH energy injections drive more
outflows in quenched than in star-forming galaxies. Both the faster
gas consumption and the feedback-driven outflows lead to lower cold
gas fraction for quenched galaxies throughout their lifetime.

Once the SMBHs seed, the inflow rates are not so distinguishable
between the populations until the quiescent galaxies quench and their
intake of cold gas decreases even more. We attribute this small drop
in the mass inflow rate to the kinetic energy released by the central
SMBH. In principle, the kinetic energy released in the radiatively
inefficient regime has both a preventative and ejective effect: it
efficiently prevents gas inflows from reaching the ISM and ejects gas
far from the ISM \citep[e.g.][]{Ramesh_2023}, so that there is no fuel
for star formation. While there is a small drop in the inflow rate,
and even a slight decrease in the outflow rate after the time of
quiescence, the cold gas fraction remains lower in quenched galaxies
than in star-forming ones at any time. This shows that the gas which
fuels star formation lies between the galaxies' ISM and the outer
limit we set for the subhalo ($R_{\rm galaxy} = 7 R_{\rm eff}$), where
we measure the mass flow rates. Therefore, AGN feedback likely
provides the pressure support for preventing gas from falling further
into the galaxies, while also ejecting gas from the ISM. We note that
the stellar feedback in the galaxies is negligible at this time, as
there is a low number of supernova explosions when the star formation
rate is so low. 

To conclude this section, quenched galaxies have lower cold gas
fractions and overall higher mass outflow rates, as the combined
effect of gas depleted to form stars and to feed the SMBH, and of
feedback-driven outflows. Mass inflow rates are higher before SMBHs
seed because quenched galaxies tend to reside in more massive halos,
and they are slightly suppressed when the galaxies become quenched as
the result from AGN kinetic energy deposition.

\subsection{Galaxy quenching in simulations} \label{subsec:comparison}

In this work, we find that IllustrisTNG predicts AGN feedback to be
the main agent of quenching, whereas environmental conditions are a
contributing factor and mergers do not seem to play any direct role.
We now compare our results with those from the literature.

Similar to our work, \citet{Hartley_2023} investigate massive galaxy
quiescence at $z\geq3$ in the TNG300 cosmological simulation. More
specifically, they study the time evolution of the first five quenched
galaxies that emerge at $z=4.2$. They conclude that AGN feedback has
an important contribution to galaxy quenching, especially when enough
kinetic energy is released by the AGN. Similarly, we find that the
suppression of star formation happens soon after AGN feedback is
triggered in the kinetic jet mode. By construction of the AGN feedback
model in IllustrisTNG, turning on the jet mode is only possible when
the SMBH reaches a certain mass and when the gas fraction is low
enough to prevent further SMBH accretion. This way, the Eddington
ratio becomes lower than the empirical quasar-to-jet mode threshold
$\chi$ (see \S~\ref{subsec:black_hole}). Because the latter scales
with the black hole mass squared, this justifies why quenching is
favored for fast SMBH growth or, alternatively, for massive SMBHs in
the IllustrisTNG simulations
\citep{Weinberger_2016,Terrazas_2020,Quai_2021,Park_2022}. Other
studies also confirm the importance of the low accretion, radiatively
inefficient AGN feedback mode in the quenching process in all the
IllustrisTNG simulations
\citep{Weinberger_2018,Terrazas_2020,Zinger_2020,Park_2022,Park_2023},
in Simba \citep{Dave_2019}, as well as in idealized simulations which
use the SWIFT code \citep{Husko_2023}. Regardless of the black hole
regime, AGN feedback is unambiguously recognized as responsible for
galaxy quenching at low redshifts (e.g.
\citealp{Dubois_2016,Beckmann_2017} in Horizon-AGN and
\citealp{Trayford_2016} in EAGLE), and our study indicates that this
conclusion is also valid at $z\geq3$ \citep[see
also][]{Lovell_2023,Lagos_2023}.

What remains more debatable is the role of the environment and of the
galaxy merger histories. In this work, we find that large-scale
environmental conditions play a minor role in quenching, mostly by
contributing to stronger accretion rates of dark matter and gas. This
1) impacts black hole seeding, which happens when galaxies are massive
enough in the IllustrisTNG simulations, and 2) provides enough gas for
the SMBHs to quickly grow and release greater amounts of energy that
can in turn impact galaxy growth. The sub-dominance of the environment
in quenching galaxies is also one the main results found by
\citet{Donnari_2021a}: as long as a galaxy is hosted in a massive
halo, internal processes such as AGN feedback are responsible for
quenching, regardless of the galaxy being a central or satellite. This
result, found by analyzing galaxies from TNG100 and TNG300, is
mitigated in the smaller volume of the simulation suite.
\citet{Park_2022} studied $z=0$ galaxies with stellar masses
$10^{10.5}-10^{11.5}\,M_\odot$ in TNG50 and found that galaxy
quenching occurs more rapidly for satellites; a similar finding is
stated by \citet{Trayford_2016} for the EAGLE simulation. The primary
reason is that satellite galaxies are more easily subject to
environmental effects, such as ram-pressure stripping and
strangulation, that can remove their star-forming gas. We note,
however, that our samples of galaxies only contain a few satellites
and that the impact of ram-pressure stripping might be more relevant
for galaxies less massive than those from our samples
\citep{Peng_2015}, which we do not investigate in this paper.
Eventually, galaxy mergers are also considered as a channel for galaxy
quenching \citep{Man_2018}. After one or several merger events,
galaxies can experience a starburst phase before star formation
dramatically decreases and all of the gas is consumed. Interestingly,
\citet{Dubois_2016} note that mergers also induce a peak of AGN
activity, which helps to quench galaxies after the burst of star
formation. Conversely, when they do not include AGN feedback in the
Horizon simulation, the supernova feedback induced after the starburst
phase is not sufficient to quench the galaxies. In TNG100, and also at
$z\leq2$, \citet{Park_2023} show that gas-rich major mergers
contribute to galaxy quenching by bringing gas inflows that feed the
central SMBH and enhancing its activity. \citet{Curro_2019}, however,
do not find a clear correlation between merger-driven starbursts and
galaxy quenching in the Simba simulation, and insist on the role of
jet mode AGN feedback instead. In our work, we find very similar
merger histories for our quiescent and star-forming samples of
galaxies. 

Therefore, we do not expect mergers to facilitate quenching at
$z\geq3$ in TNG100 and TNG300. While the quantitative effect of the
environment and of mergers may vary with time and with simulations,
the importance of AGN feedback in shutting down star formation in
massive galaxies remains to date a consensus in modern simulations,
not only at low but also at high redshifts.

\subsection{Caveats and limitations} \label{subsec:caveat}

To study massive galaxy quenching at high-redshift, we rely on
galaxies from the IllustrisTNG simulations at $z\leq3.7$. At this
redshift, we consider galaxies as quenched if their sSFR has dropped
below $10^{-10} \: \mathrm{yr}^{-1}$ \citep{Franx_2008}. In addition,
we measure the sSFR within twice the galaxies' stellar half-mass
radius, which is closer to what is done in observational works rather
than, for example, taking into account all bound stellar
particles.\footnote{25\% of the TNG100-1 $z=3.7$ quiescent sample and
52.3\% of the TNG300-1 $z=3.7$ quiescent sample are no longer
quiescent when considering all bound particles to determine the sSFR.}
As a word of caution, we note that different methods exist to classify
populations of observed galaxies (such as using a different sSFR
threshold or color-color diagrams) and may lead to different samples
of quenched galaxies \citep[see e.g.][with the IllustrisTNG
simulations at $z=0-2$]{Donnari_2019,Donnari_2021b}. 

While there is a growing number of observations showing evidence for
galaxy quenching earlier than our target redshift
\citep[e.g.][]{Girelli_2019,Nanayakkara_2022,Weaver2022,Carnall_2023Nature,Gould_2023,Valentino_2023,Antwi-Danso_2023b},
no quiescent galaxies can be found in the TNG100-1 volume beyond
$z=4.1$ using this definition. The situation gets slightly better in
TNG300-1, within which one quiescent galaxy can be found as early as
$z=4.4$. However, this remains a poor statistical sample with less
galaxies than expected when compared to the number density of massive
quiescent galaxies at $z=3-4$ \citep{Valentino_2023}. With JWST, new
observations keep pushing the detection of massive quiescent galaxies
toward higher redshifts \citep{Carnall_2023Nature}, which poses a
strong challenge to all current cosmological simulations to date since
they fail in reproducing such rare but existing objects. Therefore,
investigating the causes of quenching in the early Universe is
currently limited by the ability of numerical simulations to reproduce
quenched galaxies at high-redshift. For the last decades, cosmological
simulations have been primarily and successfully calibrated to
reproduce the properties of local galaxies. In order to make them
better suited to study the complexity of galaxy evolution at early
cosmic times, a better understanding and modeling of internal physical
processes will be crucial. 

In particular, galaxy quenching in simulations is highly related to
how black hole growth and feedback are modeled. In the IllustrisTNG
simulations, black hole seeding happens above a certain halo mass
threshold, and both the seed mass and the halo mass limit are
arbitrarily defined to match the $M_{\rm BH}-M_{\rm stellar}$ scaling
relation at $z=0$ from \citet{Kormendy_2013}. EAGLE
\citep{Schaye_2015}, Horizon-AGN \citep{Dubois_2016}, and Simba
\citep{Dave_2019} use different black hole seeding prescriptions, but
also calibrate their subgrid model to match similar $z=0$ scaling
relations (e.g. \citealp{Haring&Rix_2004} for Horizon-AGN and
\citealp{McConnell&Ma_2013} for EAGLE). Currently, the build-up of
this relation at high-redshift remains poorly constrained, and all the
cosmological simulations aforementioned predict different SMBH and
galaxy co-evolution through time. On the one hand, this is because
different subgrid models for supernovae and AGN feedback have
different efficiencies in regulating galaxy growth, leading to under
or over-massive SMBHs compared to their host galaxy masses
\citep{Habouzit_2022}. On the other hand, this may also be the result
of different black hole seeding and growth models. For example,
\citet{Bennett_2023} recently showed that allowing for earlier black
hole seeding and for super-Eddington accretion rates better match the
SMBH mass of the brightest high-redshift quasars. The black hole seed
mass is especially impactful when SMBH accretion scales with the black
hole mass squared, as in the Bondi formalism adopted in most
simulations. However, the Bondi formalism is not realistic for the
accretion of cold, turbulent gas, which makes up high-redshift
galaxies (for more realistic accretion models, see
\citealp{Debuhr_2011, Angles-Alcazar_2017, Soliman_2023}). Together
with the precise timing of black hole seeding, the SMBH accretion
scheme will determine the timescale for SMBH growth, and, as a
consequence, the exposure of the galaxy to AGN feedback. 

In this study, we find that the quenching of massive galaxies at high
redshifts happens in the IllustrisTNG simulations when the radiatively
inefficient AGN feedback mode is turned on. Nevertheless, the
transition between the quasar and the jet mode is poorly understood
and highly parametrized: it not only depends on the SMBH accretion
rate, but also on an empirically motivated black hole pivot mass and
Eddington ratio threshold. Although this parametrization contributes
to reproducing realistic galaxy populations at $z=0$, it may
artificially disfavour the emergence of galaxy quenching at
high-redshift. In addition, modeling the quasar mode via isotropic
injection of thermal energy has been shown to be barely efficient in
regulating star formation \citep{Sijacki_2007, Vogelsberger_2013}.
Instead, \citet{Choi_2015, Choi_2017, Choi_2018} model the broad line
absorption winds of AGN in the radiatively efficient regime, which is
particularly relevant at high redshifts, and show that this reduces
the sSFR and produces more realistic galaxy properties such as X-ray
luminosities. 

To improve our understanding of how AGN feedback affects star
formation, it is necessary to consider other simulations which use
different AGN feedback models to realize populations of galaxies.
Unlike TNG, not all models use an efficient jet mode, and instead rely
on more efficient quasar mode feedback or no jet feedback mode at all
to regulate galaxy growth, in an attempt to reproduce realistic galaxy
populations \citep{Henden_2018, Schaye_2015, Tremmel_2017}. Meanwhile,
higher-resolution simulations can better resolve the physical
processes resulting from SMBHs at ISM scales
\citep[e.g.][]{Wagner_Bicknell_2011, Wagner_2012, Mukherjee_2018,
Talbot_2022, Yang_2016, Meece_2017, Bourne_2021, Ehlert_2023}. These
simulations implement more detailed and physically-motivated
prescriptions of thermal and kinetic feedback and can provide
different conclusions that may or may not be complementary to the
subgrid models of cosmological simulations. We may consider
implementing more realistic models for AGN feedback like in zoom-in
simulations \citep{Cochrane_2023, Bourne_2017, Rennehan_2023,
Husko_2022, Talbot_2024}. We also note that in this work, we draw our
attention on AGN feedback and did not intend to separate its effect
from stellar feedback. Our analysis could also benefit from better
prescriptions for stellar feedback and multi-phase ISM modeling, as
this can affect the behavior of the central SMBH and how star
formation is suppressed \citep{Dubois_2014}. To better distinguish
their relative effects, an interesting follow-up would consist of
re-simulating the massive halos we targeted, similar to the 
\citet{Terrazas_2020} study at $z=0$, with and without the different
feedback channels. For instance, turning off the radiatively 
inefficient AGN feedback mode could confirm its leading role in 
quenching galaxies.


\section{Conclusion}  \label{sec:conclusion}

Using the suite of cosmological simulations IllustrisTNG, we performed
an analysis of quiescent, massive galaxies at $z\gtrsim3$. The goal
was to understand what suppresses their star formation early in their
evolution, and how other similar mass galaxies remain highly star
forming. We selected samples of quiescent and star-forming galaxies in
three different TNG simulation sets: galaxies from TNG100-1 at $z=3.7$
and $z=3$, and from TNG300-1 at $z=3.7$. Our main findings are
summarized below:

\begin{enumerate}

\item \textit{\textbf{There is a clear correlation between galaxy
quenching and AGN feedback}}: quenched galaxies are exposed for longer
to stronger AGN activity than star-forming galaxies. Indeed, leading
up to the time of quenching, SMBHs in quenched galaxies accrete at a
higher rate and accordingly release larger amounts of thermal energy
than those in star-forming galaxies. 

\item \textit{\textbf{Galaxies become quenched when the radiatively
inefficient kinetic AGN feedback mode is triggered.}} Because this
happens as soon as the SMBH mass and accretion rate reach a certain
threshold, quenching in the TNG simulations is favoured for galaxies
whose gas supply enables a fast black hole growth.

\item \textit{\textbf{Large-scale environmental conditions likely
contribute to quenching at high-redshift}} by enabling the fast and
sustained SMBH growth required to trigger AGN feedback in the jet
mode. The simulated quiescent galaxies in general reside in more
massive host halos and in denser environments throughout their
lifetime. As a consequence from their deeper gravitational potential
(and likely to their proximity to cosmic filaments), they accrete dark
matter and gas at a higher rate than star-forming galaxies, which
allows faster stellar mass and black hole growth. 

\item \textit{\textbf{Merger events are not important to
distinguishing between quiescent and star-forming galaxies}} at high
redshifts, and are unlikely to favour galaxy quenching at $z\geq3$.

\end{enumerate}

This works gives strong evidence that AGN feedback is the dominant
cause of quiescence at high-redshift in the IllustrisTNG simulations.
Quenching coincides with the triggering of the kinetic, jet mode AGN
feedback when black holes enter the radiatively inefficient regime.
For this reason, the precise timing of galaxy quenching is highly
dependent on the numerical modeling of AGN feedback, initially tuned
so that simulations at $z=0$ resemble their observed analogs. As a
side effect, massive galaxies do not quench before a certain time, and
cosmological simulations to-date commonly underestimate the number
density of massive quenched galaxies at $z>3$. For remedying this
issue, one interesting avenue would be to improve our AGN feedback
models, especially in the radiatively efficient regime which dominates
at early cosmic epochs. This would help a better understanding of the
role of AGN feedback in quenching galaxies, from the cosmic dawn to
present-day.

\section*{Data Availability}                                          

Data directly related to the figures of this publication are available
on request from SK. The IllustrisTNG simulations are publicly available at \url{www.tng-project.org/data} \citep{Nelson_2019}. \\

\section*{Acknowledgements}                                           

The authors would like to thank the referees for useful comments in
the revision of this manuscript. SK acknowledges support from the U.S. Fulbright Scholarship program at the time of completing this work, and
is thankful to Annalisa Pillepich for insightful conversations as well
as the opportunity to present the results at the GASPISA2024---``The
Physical Processes Shaping the Stellar and Gaseous Histories of
Galaxies''--- conference. MH and MF acknowledge funding from the Swiss
National Science Foundation (SNF) via a PRIMA Grant PR00P2 193577 “From
Cosmic Dawn to High Noon: The Role of Black Holes for Young Galaxies.” 
FV at is funded by the Danish National Research Foundation under Grant
No. 140.

\bibliographystyle{mnras} 
\typeout{}
\bibliography{refs}

\begin{thebibliography}{}
\makeatletter
\relax
\def\mn@urlcharsother{\let\do\@makeother \do\$\do\&\do\#\do\^\do\_\do\%\do\~}
\def\mn@doi{\begingroup\mn@urlcharsother \@ifnextchar [ {\mn@doi@} {\mn@doi@[]}}
\def\mn@doi@[#1]#2{\def\@tempa{#1}\ifx\@tempa\@empty \href {http://dx.doi.org/#2} {doi:#2}\else \href {http://dx.doi.org/#2} {#1}\fi \endgroup}
\def\mn@eprint#1#2{\mn@eprint@#1:#2::\@nil}
\def\mn@eprint@arXiv#1{\href {http://arxiv.org/abs/#1} {{\tt arXiv:#1}}}
\def\mn@eprint@dblp#1{\href {http://dblp.uni-trier.de/rec/bibtex/#1.xml} {dblp:#1}}
\def\mn@eprint@#1:#2:#3:#4\@nil{\def\@tempa {#1}\def\@tempb {#2}\def\@tempc {#3}\ifx \@tempc \@empty \let \@tempc \@tempb \let \@tempb \@tempa \fi \ifx \@tempb \@empty \def\@tempb {arXiv}\fi \@ifundefined {mn@eprint@\@tempb}{\@tempb:\@tempc}{\expandafter \expandafter \csname mn@eprint@\@tempb\endcsname \expandafter{\@tempc}}}

\bibitem[\protect\citeauthoryear{{Angl{\'e}s-Alc{\'a}zar}, {Dav{\'e}}, {Faucher-Gigu{\`e}re}, {{\"O}zel}  \& {Hopkins}}{{Angl{\'e}s-Alc{\'a}zar} et~al.}{2017}]{Angles-Alcazar_2017}
{Angl{\'e}s-Alc{\'a}zar} D.,  {Dav{\'e}} R.,  {Faucher-Gigu{\`e}re} C.-A.,  {{\"O}zel} F.,   {Hopkins} P.~F.,  2017, \mn@doi [\mnras] {10.1093/mnras/stw2565}, \href {https://ui.adsabs.harvard.edu/abs/2017MNRAS.464.2840A} {464, 2840}

\bibitem[\protect\citeauthoryear{{Antwi-Danso} et~al.,}{{Antwi-Danso} et~al.}{2023a}]{Antwi-Danso_2023b}
{Antwi-Danso} J.,  et~al., 2023a, \mn@doi [arXiv e-prints] {10.48550/arXiv.2307.09590}, \href {https://ui.adsabs.harvard.edu/abs/2023arXiv230709590A} {p. arXiv:2307.09590}

\bibitem[\protect\citeauthoryear{{Antwi-Danso} et~al.,}{{Antwi-Danso} et~al.}{2023b}]{Antwi-Danso_2023}
{Antwi-Danso} J.,  et~al., 2023b, \mn@doi [\apj] {10.3847/1538-4357/aca294}, \href {https://ui.adsabs.harvard.edu/abs/2023ApJ...943..166A} {943, 166}

\bibitem[\protect\citeauthoryear{{Arav}, {Borguet}, {Chamberlain}, {Edmonds}  \& {Danforth}}{{Arav} et~al.}{2013}]{Arav_2013}
{Arav} N.,  {Borguet} B.,  {Chamberlain} C.,  {Edmonds} D.,   {Danforth} C.,  2013, \mn@doi [\mnras] {10.1093/mnras/stt1812}, \href {https://ui.adsabs.harvard.edu/abs/2013MNRAS.436.3286A} {436, 3286}

\bibitem[\protect\citeauthoryear{{Beckmann} et~al.,}{{Beckmann} et~al.}{2017}]{Beckmann_2017}
{Beckmann} R.~S.,  et~al., 2017, \mn@doi [\mnras] {10.1093/mnras/stx1831}, \href {https://ui.adsabs.harvard.edu/abs/2017MNRAS.472..949B} {472, 949}

\bibitem[\protect\citeauthoryear{{Bennett}, {Sijacki}, {Costa}, {Laporte}  \& {Witten}}{{Bennett} et~al.}{2023}]{Bennett_2023}
{Bennett} J.~S.,  {Sijacki} D.,  {Costa} T.,  {Laporte} N.,   {Witten} C.,  2023, \mn@doi [arXiv e-prints] {10.48550/arXiv.2305.11932}, \href {https://ui.adsabs.harvard.edu/abs/2023arXiv230511932B} {p. arXiv:2305.11932}

\bibitem[\protect\citeauthoryear{{Bocquet}, {Saro}, {Dolag}  \& {Mohr}}{{Bocquet} et~al.}{2016}]{Bocquet_2016}
{Bocquet} S.,  {Saro} A.,  {Dolag} K.,   {Mohr} J.~J.,  2016, \mn@doi [\mnras] {10.1093/mnras/stv2657}, \href {https://ui.adsabs.harvard.edu/abs/2016MNRAS.456.2361B} {456, 2361}

\bibitem[\protect\citeauthoryear{Book \& Benson}{Book \& Benson}{2010}]{Book_2010}
Book L.~G.,  Benson A.~J.,  2010, \mn@doi [\apj] {10.1088/0004-637x/716/1/810}, 716, 810–818

\bibitem[\protect\citeauthoryear{Boselli et~al.,}{Boselli et~al.}{2016}]{Boselli_2016}
Boselli A.,  et~al., 2016, \mn@doi [\aap] {10.1051/0004-6361/201629221}, 596, A11

\bibitem[\protect\citeauthoryear{{Bourne} \& {Sijacki}}{{Bourne} \& {Sijacki}}{2017}]{Bourne_2017}
{Bourne} M.~A.,  {Sijacki} D.,  2017, \mn@doi [\mnras] {10.1093/mnras/stx2269}, \href {https://ui.adsabs.harvard.edu/abs/2017MNRAS.472.4707B} {472, 4707}

\bibitem[\protect\citeauthoryear{{Bourne} \& {Sijacki}}{{Bourne} \& {Sijacki}}{2021}]{Bourne_2021}
{Bourne} M.~A.,  {Sijacki} D.,  2021, \mn@doi [\mnras] {10.1093/mnras/stab1662}, \href {https://ui.adsabs.harvard.edu/abs/2021MNRAS.506..488B} {506, 488}

\bibitem[\protect\citeauthoryear{{Bower}, {Benson}, {Malbon}, {Helly}, {Frenk}, {Baugh}, {Cole}  \& {Lacey}}{{Bower} et~al.}{2006}]{Bower_2006}
{Bower} R.~G.,  {Benson} A.~J.,  {Malbon} R.,  {Helly} J.~C.,  {Frenk} C.~S.,  {Baugh} C.~M.,  {Cole} S.,   {Lacey} C.~G.,  2006, \mn@doi [\mnras] {10.1111/j.1365-2966.2006.10519.x}, \href {https://ui.adsabs.harvard.edu/abs/2006MNRAS.370..645B} {370, 645}

\bibitem[\protect\citeauthoryear{{Cano-D{\'\i}az}, {Maiolino}, {Marconi}, {Netzer}, {Shemmer}  \& {Cresci}}{{Cano-D{\'\i}az} et~al.}{2012}]{Cano-Diaz_2012}
{Cano-D{\'\i}az} M.,  {Maiolino} R.,  {Marconi} A.,  {Netzer} H.,  {Shemmer} O.,   {Cresci} G.,  2012, \mn@doi [\aap] {10.1051/0004-6361/201118358}, \href {https://ui.adsabs.harvard.edu/abs/2012A&A...537L...8C} {537, L8}

\bibitem[\protect\citeauthoryear{{Carnall} et~al.,}{{Carnall} et~al.}{2023a}]{Carnall_2023}
{Carnall} A.~C.,  et~al., 2023a, \mn@doi [\mnras] {10.1093/mnras/stad369}, \href {https://ui.adsabs.harvard.edu/abs/2023MNRAS.520.3974C} {520, 3974}

\bibitem[\protect\citeauthoryear{{Carnall} et~al.,}{{Carnall} et~al.}{2023b}]{Carnall_2023Nature}
{Carnall} A.~C.,  et~al., 2023b, \mn@doi [\nat] {10.1038/s41586-023-06158-6}, \href {https://ui.adsabs.harvard.edu/abs/2023Natur.619..716C} {619, 716}

\bibitem[\protect\citeauthoryear{{Cecchi}, {Bolzonella}, {Cimatti}  \& {Girelli}}{{Cecchi} et~al.}{2019}]{Cecchi_2019}
{Cecchi} R.,  {Bolzonella} M.,  {Cimatti} A.,   {Girelli} G.,  2019, \mn@doi [\apjl] {10.3847/2041-8213/ab2c80}, \href {https://ui.adsabs.harvard.edu/abs/2019ApJ...880L..14C} {880, L14}

\bibitem[\protect\citeauthoryear{{Chevalier} \& {Clegg}}{{Chevalier} \& {Clegg}}{1985}]{Chevalier&Clegg1985}
{Chevalier} R.~A.,  {Clegg} A.~W.,  1985, \mn@doi [\nat] {10.1038/317044a0}, \href {https://ui.adsabs.harvard.edu/abs/1985Natur.317...44C} {317, 44}

\bibitem[\protect\citeauthoryear{{Choi}, {Ostriker}, {Naab}, {Oser}  \& {Moster}}{{Choi} et~al.}{2015}]{Choi_2015}
{Choi} E.,  {Ostriker} J.~P.,  {Naab} T.,  {Oser} L.,   {Moster} B.~P.,  2015, \mn@doi [\mnras] {10.1093/mnras/stv575}, \href {https://ui.adsabs.harvard.edu/abs/2015MNRAS.449.4105C} {449, 4105}

\bibitem[\protect\citeauthoryear{{Choi}, {Ostriker}, {Naab}, {Somerville}, {Hirschmann}, {N{\'u}{\~n}ez}, {Hu}  \& {Oser}}{{Choi} et~al.}{2017}]{Choi_2017}
{Choi} E.,  {Ostriker} J.~P.,  {Naab} T.,  {Somerville} R.~S.,  {Hirschmann} M.,  {N{\'u}{\~n}ez} A.,  {Hu} C.-Y.,   {Oser} L.,  2017, \mn@doi [\apj] {10.3847/1538-4357/aa7849}, \href {https://ui.adsabs.harvard.edu/abs/2017ApJ...844...31C} {844, 31}

\bibitem[\protect\citeauthoryear{{Choi}, {Somerville}, {Ostriker}, {Naab}  \& {Hirschmann}}{{Choi} et~al.}{2018}]{Choi_2018}
{Choi} E.,  {Somerville} R.~S.,  {Ostriker} J.~P.,  {Naab} T.,   {Hirschmann} M.,  2018, \mn@doi [\apj] {10.3847/1538-4357/aae076}, \href {https://ui.adsabs.harvard.edu/abs/2018ApJ...866...91C} {866, 91}

\bibitem[\protect\citeauthoryear{{Cicone} et~al.,}{{Cicone} et~al.}{2014}]{Cicone2014}
{Cicone} C.,  et~al., 2014, \mn@doi [\aap] {10.1051/0004-6361/201322464}, \href {https://ui.adsabs.harvard.edu/abs/2014A&A...562A..21C} {562, A21}

\bibitem[\protect\citeauthoryear{{Ciotti}, {D'Ercole}, {Pellegrini}  \& {Renzini}}{{Ciotti} et~al.}{1991}]{Ciotti_1991}
{Ciotti} L.,  {D'Ercole} A.,  {Pellegrini} S.,   {Renzini} A.,  1991, \mn@doi [\apj] {10.1086/170289}, \href {https://ui.adsabs.harvard.edu/abs/1991ApJ...376..380C} {376, 380}

\bibitem[\protect\citeauthoryear{{Cochrane} et~al.,}{{Cochrane} et~al.}{2023}]{Cochrane_2023}
{Cochrane} R.~K.,  et~al., 2023, \mn@doi [\mnras] {10.1093/mnras/stad1528}, \href {https://ui.adsabs.harvard.edu/abs/2023MNRAS.523.2409C} {523, 2409}

\bibitem[\protect\citeauthoryear{{Costa}, {Rosdahl}, {Sijacki}  \& {Haehnelt}}{{Costa} et~al.}{2018}]{Costa_2018}
{Costa} T.,  {Rosdahl} J.,  {Sijacki} D.,   {Haehnelt} M.~G.,  2018, \mn@doi [\mnras] {10.1093/mnras/sty1514}, \href {https://ui.adsabs.harvard.edu/abs/2018MNRAS.479.2079C} {479, 2079}

\bibitem[\protect\citeauthoryear{{Croton} et~al.,}{{Croton} et~al.}{2006}]{Croton_2006}
{Croton} D.~J.,  et~al., 2006, \mn@doi [\mnras] {10.1111/j.1365-2966.2005.09675.x}, \href {https://ui.adsabs.harvard.edu/abs/2006MNRAS.365...11C} {365, 11}

\bibitem[\protect\citeauthoryear{{Dav{\'e}}, {Angl{\'e}s-Alc{\'a}zar}, {Narayanan}, {Li}, {Rafieferantsoa}  \& {Appleby}}{{Dav{\'e}} et~al.}{2019}]{Dave_2019}
{Dav{\'e}} R.,  {Angl{\'e}s-Alc{\'a}zar} D.,  {Narayanan} D.,  {Li} Q.,  {Rafieferantsoa} M.~H.,   {Appleby} S.,  2019, \mn@doi [\mnras] {10.1093/mnras/stz937}, \href {https://ui.adsabs.harvard.edu/abs/2019MNRAS.486.2827D} {486, 2827}

\bibitem[\protect\citeauthoryear{{Debuhr}, {Quataert}  \& {Ma}}{{Debuhr} et~al.}{2011}]{Debuhr_2011}
{Debuhr} J.,  {Quataert} E.,   {Ma} C.-P.,  2011, \mn@doi [\mnras] {10.1111/j.1365-2966.2010.17992.x}, \href {https://ui.adsabs.harvard.edu/abs/2011MNRAS.412.1341D} {412, 1341}

\bibitem[\protect\citeauthoryear{{Di Matteo}, {Springel}  \& {Hernquist}}{{Di Matteo} et~al.}{2005}]{DiMatteo_2005}
{Di Matteo} T.,  {Springel} V.,   {Hernquist} L.,  2005, \mn@doi [\nat] {10.1038/nature03335}, \href {https://ui.adsabs.harvard.edu/abs/2005Natur.433..604D} {433, 604}

\bibitem[\protect\citeauthoryear{{Donnari} et~al.,}{{Donnari} et~al.}{2019}]{Donnari_2019}
{Donnari} M.,  et~al., 2019, \mn@doi [\mnras] {10.1093/mnras/stz712}, \href {https://ui.adsabs.harvard.edu/abs/2019MNRAS.485.4817D} {485, 4817}

\bibitem[\protect\citeauthoryear{{Donnari} et~al.,}{{Donnari} et~al.}{2021a}]{Donnari_2021a}
{Donnari} M.,  et~al., 2021a, \mn@doi [\mnras] {10.1093/mnras/staa3006}, \href {https://ui.adsabs.harvard.edu/abs/2021MNRAS.500.4004D} {500, 4004}

\bibitem[\protect\citeauthoryear{{Donnari}, {Pillepich}, {Nelson}, {Marinacci}, {Vogelsberger}  \& {Hernquist}}{{Donnari} et~al.}{2021b}]{Donnari_2021b}
{Donnari} M.,  {Pillepich} A.,  {Nelson} D.,  {Marinacci} F.,  {Vogelsberger} M.,   {Hernquist} L.,  2021b, \mn@doi [\mnras] {10.1093/mnras/stab1950}, \href {https://ui.adsabs.harvard.edu/abs/2021MNRAS.506.4760D} {506, 4760}

\bibitem[\protect\citeauthoryear{{Dubois}, {Devriendt}, {Slyz}  \& {Teyssier}}{{Dubois} et~al.}{2012}]{Dubois_2012}
{Dubois} Y.,  {Devriendt} J.,  {Slyz} A.,   {Teyssier} R.,  2012, \mn@doi [\mnras] {10.1111/j.1365-2966.2011.20236.x}, \href {https://ui.adsabs.harvard.edu/abs/2012MNRAS.420.2662D} {420, 2662}

\bibitem[\protect\citeauthoryear{{Dubois}, {Volonteri}, {Silk}, {Devriendt}  \& {Slyz}}{{Dubois} et~al.}{2014}]{Dubois_2014}
{Dubois} Y.,  {Volonteri} M.,  {Silk} J.,  {Devriendt} J.,   {Slyz} A.,  2014, \mn@doi [\mnras] {10.1093/mnras/stu425}, \href {https://ui.adsabs.harvard.edu/abs/2014MNRAS.440.2333D} {440, 2333}

\bibitem[\protect\citeauthoryear{{Dubois}, {Peirani}, {Pichon}, {Devriendt}, {Gavazzi}, {Welker}  \& {Volonteri}}{{Dubois} et~al.}{2016}]{Dubois_2016}
{Dubois} Y.,  {Peirani} S.,  {Pichon} C.,  {Devriendt} J.,  {Gavazzi} R.,  {Welker} C.,   {Volonteri} M.,  2016, \mn@doi [\mnras] {10.1093/mnras/stw2265}, \href {https://ui.adsabs.harvard.edu/abs/2016MNRAS.463.3948D} {463, 3948}

\bibitem[\protect\citeauthoryear{{Ehlert}, {Weinberger}, {Pfrommer}, {Pakmor}  \& {Springel}}{{Ehlert} et~al.}{2023}]{Ehlert_2023}
{Ehlert} K.,  {Weinberger} R.,  {Pfrommer} C.,  {Pakmor} R.,   {Springel} V.,  2023, \mn@doi [\mnras] {10.1093/mnras/stac2860}, \href {https://ui.adsabs.harvard.edu/abs/2023MNRAS.518.4622E} {518, 4622}

\bibitem[\protect\citeauthoryear{{Fabian}}{{Fabian}}{2012}]{Fabian_2012}
{Fabian} A.~C.,  2012, \mn@doi [\araa] {10.1146/annurev-astro-081811-125521}, \href {https://ui.adsabs.harvard.edu/abs/2012ARA&A..50..455F} {50, 455}

\bibitem[\protect\citeauthoryear{{Faucher-Gigu{\`e}re} \& {Quataert}}{{Faucher-Gigu{\`e}re} \& {Quataert}}{2012}]{Faucher-Giguere_2012}
{Faucher-Gigu{\`e}re} C.-A.,  {Quataert} E.,  2012, \mn@doi [\mnras] {10.1111/j.1365-2966.2012.21512.x}, \href {https://ui.adsabs.harvard.edu/abs/2012MNRAS.425..605F} {425, 605}

\bibitem[\protect\citeauthoryear{{Feruglio}, {Maiolino}, {Piconcelli}, {Menci}, {Aussel}, {Lamastra}  \& {Fiore}}{{Feruglio} et~al.}{2010}]{Feruglio_2010}
{Feruglio} C.,  {Maiolino} R.,  {Piconcelli} E.,  {Menci} N.,  {Aussel} H.,  {Lamastra} A.,   {Fiore} F.,  2010, \mn@doi [\aap] {10.1051/0004-6361/201015164}, \href {https://ui.adsabs.harvard.edu/abs/2010A&A...518L.155F} {518, L155}

\bibitem[\protect\citeauthoryear{{Fischer} et~al.,}{{Fischer} et~al.}{2010}]{Fischer_2010}
{Fischer} J.,  et~al., 2010, \mn@doi [\aap] {10.1051/0004-6361/201014676}, \href {https://ui.adsabs.harvard.edu/abs/2010A&A...518L..41F} {518, L41}

\bibitem[\protect\citeauthoryear{{Forrest} et~al.,}{{Forrest} et~al.}{2020}]{Forrest_2020}
{Forrest} B.,  et~al., 2020, \mn@doi [\apj] {10.3847/1538-4357/abb819}, \href {https://ui.adsabs.harvard.edu/abs/2020ApJ...903...47F} {903, 47}

\bibitem[\protect\citeauthoryear{{F{\"o}rster Schreiber} et~al.,}{{F{\"o}rster Schreiber} et~al.}{2014}]{ForsterSchreiber2014}
{F{\"o}rster Schreiber} N.~M.,  et~al., 2014, \mn@doi [\apj] {10.1088/0004-637X/787/1/38}, \href {https://ui.adsabs.harvard.edu/abs/2014ApJ...787...38F} {787, 38}

\bibitem[\protect\citeauthoryear{{Franx}, {van Dokkum}, {F{\"o}rster Schreiber}, {Wuyts}, {Labb{\'e}}  \& {Toft}}{{Franx} et~al.}{2008}]{Franx_2008}
{Franx} M.,  {van Dokkum} P.~G.,  {F{\"o}rster Schreiber} N.~M.,  {Wuyts} S.,  {Labb{\'e}} I.,   {Toft} S.,  2008, \mn@doi [\apj] {10.1086/592431}, \href {https://ui.adsabs.harvard.edu/abs/2008ApJ...688..770F} {688, 770}

\bibitem[\protect\citeauthoryear{Genel et~al.,}{Genel et~al.}{2014}]{Genel_2014}
Genel S.,  et~al., 2014, \mn@doi [\mnras] {10.1093/mnras/stu1654}, 445, 175

\bibitem[\protect\citeauthoryear{{Genzel} et~al.,}{{Genzel} et~al.}{2014}]{Genzel2014}
{Genzel} R.,  et~al., 2014, \mn@doi [\apj] {10.1088/0004-637X/796/1/7}, \href {https://ui.adsabs.harvard.edu/abs/2014ApJ...796....7G} {796, 7}

\bibitem[\protect\citeauthoryear{{Girelli}, {Bolzonella}  \& {Cimatti}}{{Girelli} et~al.}{2019}]{Girelli_2019}
{Girelli} G.,  {Bolzonella} M.,   {Cimatti} A.,  2019, \mn@doi [\aap] {10.1051/0004-6361/201834547}, \href {https://ui.adsabs.harvard.edu/abs/2019A&A...632A..80G} {632, A80}

\bibitem[\protect\citeauthoryear{{Gould} et~al.,}{{Gould} et~al.}{2023}]{Gould_2023}
{Gould} K. M.~L.,  et~al., 2023, \mn@doi [\aj] {10.3847/1538-3881/accadc}, \href {https://ui.adsabs.harvard.edu/abs/2023AJ....165..248G} {165, 248}

\bibitem[\protect\citeauthoryear{{Habouzit} et~al.,}{{Habouzit} et~al.}{2022}]{Habouzit_2022}
{Habouzit} M.,  et~al., 2022, \mn@doi [\mnras] {10.1093/mnras/stac225}, \href {https://ui.adsabs.harvard.edu/abs/2022MNRAS.511.3751H} {511, 3751}

\bibitem[\protect\citeauthoryear{{Hardcastle}}{{Hardcastle}}{2018}]{Hardcastle_2018}
{Hardcastle} M.,  2018, \mn@doi [Nature Astronomy] {10.1038/s41550-018-0424-1}, \href {https://ui.adsabs.harvard.edu/abs/2018NatAs...2..273H} {2, 273}

\bibitem[\protect\citeauthoryear{{Hardcastle} \& {Croston}}{{Hardcastle} \& {Croston}}{2020}]{Harcastle_Croton_2020}
{Hardcastle} M.~J.,  {Croston} J.~H.,  2020, \mn@doi [\nar] {10.1016/j.newar.2020.101539}, \href {https://ui.adsabs.harvard.edu/abs/2020NewAR..8801539H} {88, 101539}

\bibitem[\protect\citeauthoryear{{H{\"a}ring} \& {Rix}}{{H{\"a}ring} \& {Rix}}{2004}]{Haring&Rix_2004}
{H{\"a}ring} N.,  {Rix} H.-W.,  2004, \mn@doi [\apjl] {10.1086/383567}, \href {https://ui.adsabs.harvard.edu/abs/2004ApJ...604L..89H} {604, L89}

\bibitem[\protect\citeauthoryear{{Harrison}, {Alexander}, {Mullaney}  \& {Swinbank}}{{Harrison} et~al.}{2014}]{Harrison_2014}
{Harrison} C.~M.,  {Alexander} D.~M.,  {Mullaney} J.~R.,   {Swinbank} A.~M.,  2014, \mn@doi [\mnras] {10.1093/mnras/stu515}, \href {https://ui.adsabs.harvard.edu/abs/2014MNRAS.441.3306H} {441, 3306}

\bibitem[\protect\citeauthoryear{Harrison et~al.,}{Harrison et~al.}{2015}]{Harrison_2015}
Harrison C.~M.,  et~al., 2015, \mn@doi [\mnras] {10.1093/mnras/stv2727}, 456, 1195

\bibitem[\protect\citeauthoryear{{Hartley} et~al.,}{{Hartley} et~al.}{2023}]{Hartley_2023}
{Hartley} A.~I.,  et~al., 2023, \mn@doi [\mnras] {10.1093/mnras/stad1162}, \href {https://ui.adsabs.harvard.edu/abs/2023MNRAS.522.3138H} {522, 3138}

\bibitem[\protect\citeauthoryear{{Henden}, {Puchwein}, {Shen}  \& {Sijacki}}{{Henden} et~al.}{2018}]{Henden_2018}
{Henden} N.~A.,  {Puchwein} E.,  {Shen} S.,   {Sijacki} D.,  2018, \mn@doi [\mnras] {10.1093/mnras/sty1780}, \href {https://ui.adsabs.harvard.edu/abs/2018MNRAS.479.5385H} {479, 5385}

\bibitem[\protect\citeauthoryear{{Hirschmann}, {Dolag}, {Saro}, {Bachmann}, {Borgani}  \& {Burkert}}{{Hirschmann} et~al.}{2014a}]{Hirschmann_2014}
{Hirschmann} M.,  {Dolag} K.,  {Saro} A.,  {Bachmann} L.,  {Borgani} S.,   {Burkert} A.,  2014a, \mn@doi [\mnras] {10.1093/mnras/stu1023}, \href {https://ui.adsabs.harvard.edu/abs/2014MNRAS.442.2304H} {442, 2304}

\bibitem[\protect\citeauthoryear{{Hirschmann}, {De Lucia}, {Wilman}, {Weinmann}, {Iovino}, {Cucciati}, {Zibetti}  \& {Villalobos}}{{Hirschmann} et~al.}{2014b}]{Hirschmann_2014b}
{Hirschmann} M.,  {De Lucia} G.,  {Wilman} D.,  {Weinmann} S.,  {Iovino} A.,  {Cucciati} O.,  {Zibetti} S.,   {Villalobos} {\'A}.,  2014b, \mn@doi [\mnras] {10.1093/mnras/stu1609}, \href {https://ui.adsabs.harvard.edu/abs/2014MNRAS.444.2938H} {444, 2938}

\bibitem[\protect\citeauthoryear{{Hu{\v{s}}ko}, {Lacey}, {Schaye}, {Schaller}  \& {Nobels}}{{Hu{\v{s}}ko} et~al.}{2022}]{Husko_2022}
{Hu{\v{s}}ko} F.,  {Lacey} C.~G.,  {Schaye} J.,  {Schaller} M.,   {Nobels} F. S.~J.,  2022, \mn@doi [\mnras] {10.1093/mnras/stac2278}, \href {https://ui.adsabs.harvard.edu/abs/2022MNRAS.516.3750H} {516, 3750}

\bibitem[\protect\citeauthoryear{{Hu{\v{s}}ko}, {Lacey}, {Schaye}, {Nobels}  \& {Schaller}}{{Hu{\v{s}}ko} et~al.}{2023}]{Husko_2023}
{Hu{\v{s}}ko} F.,  {Lacey} C.~G.,  {Schaye} J.,  {Nobels} F. S.~J.,   {Schaller} M.,  2023, \mn@doi [arXiv e-prints] {10.48550/arXiv.2307.01409}, \href {https://ui.adsabs.harvard.edu/abs/2023arXiv230701409H} {p. arXiv:2307.01409}

\bibitem[\protect\citeauthoryear{{Ishibashi} \& {Fabian}}{{Ishibashi} \& {Fabian}}{2012}]{Ishibashi_2012}
{Ishibashi} W.,  {Fabian} A.~C.,  2012, \mn@doi [\mnras] {10.1111/j.1365-2966.2012.22074.x}, \href {https://ui.adsabs.harvard.edu/abs/2012MNRAS.427.2998I} {427, 2998}

\bibitem[\protect\citeauthoryear{{Ishibashi} \& {Fabian}}{{Ishibashi} \& {Fabian}}{2015}]{Ishibashi_2015}
{Ishibashi} W.,  {Fabian} A.~C.,  2015, \mn@doi [\mnras] {10.1093/mnras/stv944}, \href {https://ui.adsabs.harvard.edu/abs/2015MNRAS.451...93I} {451, 93}

\bibitem[\protect\citeauthoryear{{Khochfar} \& {Ostriker}}{{Khochfar} \& {Ostriker}}{2008}]{Khochfar_2008}
{Khochfar} S.,  {Ostriker} J.~P.,  2008, \mn@doi [\apj] {10.1086/587470}, \href {https://ui.adsabs.harvard.edu/abs/2008ApJ...680...54K} {680, 54}

\bibitem[\protect\citeauthoryear{{King} \& {Pounds}}{{King} \& {Pounds}}{2015}]{King_Pounds_2015}
{King} A.,  {Pounds} K.,  2015, \mn@doi [\araa] {10.1146/annurev-astro-082214-122316}, \href {https://ui.adsabs.harvard.edu/abs/2015ARA&A..53..115K} {53, 115}

\bibitem[\protect\citeauthoryear{{Kormendy} \& {Ho}}{{Kormendy} \& {Ho}}{2013}]{Kormendy_2013}
{Kormendy} J.,  {Ho} L.~C.,  2013, \mn@doi [\araa] {10.1146/annurev-astro-082708-101811}, \href {https://ui.adsabs.harvard.edu/abs/2013ARA&A..51..511K} {51, 511}

\bibitem[\protect\citeauthoryear{{Kravtsov}}{{Kravtsov}}{2013}]{Kravtsov_2013}
{Kravtsov} A.~V.,  2013, \mn@doi [\apjl] {10.1088/2041-8205/764/2/L31}, \href {https://ui.adsabs.harvard.edu/abs/2013ApJ...764L..31K} {764, L31}

\bibitem[\protect\citeauthoryear{{Lagos} et~al.,}{{Lagos} et~al.}{2023}]{Lagos_2023}
{Lagos} C. D.~P.,  et~al., 2023, \mn@doi [arXiv e-prints] {10.48550/arXiv.2309.02310}, \href {https://ui.adsabs.harvard.edu/abs/2023arXiv230902310L} {p. arXiv:2309.02310}

\bibitem[\protect\citeauthoryear{{Long} et~al.,}{{Long} et~al.}{2023}]{Long_2023}
{Long} A.~S.,  et~al., 2023, \mn@doi [arXiv e-prints] {10.48550/arXiv.2305.04662}, \href {https://ui.adsabs.harvard.edu/abs/2023arXiv230504662L} {p. arXiv:2305.04662}

\bibitem[\protect\citeauthoryear{{Lovell} et~al.,}{{Lovell} et~al.}{2023}]{Lovell_2023}
{Lovell} C.~C.,  et~al., 2023, \mn@doi [\mnras] {10.1093/mnras/stad2550}, \href {https://ui.adsabs.harvard.edu/abs/2023MNRAS.525.5520L} {525, 5520}

\bibitem[\protect\citeauthoryear{Man \& Belli}{Man \& Belli}{2018}]{Man_2018}
Man A.,  Belli S.,  2018, \mn@doi [Nature Astronomy] {10.1038/s41550-018-0558-1}, 2, 695

\bibitem[\protect\citeauthoryear{{Marinacci} et~al.,}{{Marinacci} et~al.}{2018}]{Marinacci_2018}
{Marinacci} F.,  et~al., 2018, \mn@doi [\mnras] {10.1093/mnras/sty2206}, \href {https://ui.adsabs.harvard.edu/abs/2018MNRAS.480.5113M} {480, 5113}

\bibitem[\protect\citeauthoryear{{McConnell} \& {Ma}}{{McConnell} \& {Ma}}{2013}]{McConnell&Ma_2013}
{McConnell} N.~J.,  {Ma} C.-P.,  2013, \mn@doi [\apj] {10.1088/0004-637X/764/2/184}, \href {https://ui.adsabs.harvard.edu/abs/2013ApJ...764..184M} {764, 184}

\bibitem[\protect\citeauthoryear{{Meece}, {Voit}  \& {O'Shea}}{{Meece} et~al.}{2017}]{Meece_2017}
{Meece} G.~R.,  {Voit} G.~M.,   {O'Shea} B.~W.,  2017, \mn@doi [\apj] {10.3847/1538-4357/aa6fb1}, \href {https://ui.adsabs.harvard.edu/abs/2017ApJ...841..133M} {841, 133}

\bibitem[\protect\citeauthoryear{Merlin et~al.,}{Merlin et~al.}{2019}]{Merlin_2019}
Merlin E.,  et~al., 2019, \mn@doi [\mnras] {10.1093/mnras/stz2615}, 490, 3309–3328

\bibitem[\protect\citeauthoryear{{Mihos} \& {Hernquist}}{{Mihos} \& {Hernquist}}{1996}]{Mihos_Hernquist_1996}
{Mihos} J.~C.,  {Hernquist} L.,  1996, \mn@doi [\apj] {10.1086/177353}, \href {https://ui.adsabs.harvard.edu/abs/1996ApJ...464..641M} {464, 641}

\bibitem[\protect\citeauthoryear{{Mukherjee}, {Bicknell}, {Wagner}, {Sutherland}  \& {Silk}}{{Mukherjee} et~al.}{2018}]{Mukherjee_2018}
{Mukherjee} D.,  {Bicknell} G.~V.,  {Wagner} A.~Y.,  {Sutherland} R.~S.,   {Silk} J.,  2018, \mn@doi [\mnras] {10.1093/mnras/sty1776}, \href {https://ui.adsabs.harvard.edu/abs/2018MNRAS.479.5544M} {479, 5544}

\bibitem[\protect\citeauthoryear{{Naiman} et~al.,}{{Naiman} et~al.}{2018}]{Naiman_2018}
{Naiman} J.~P.,  et~al., 2018, \mn@doi [\mnras] {10.1093/mnras/sty618}, \href {https://ui.adsabs.harvard.edu/abs/2018MNRAS.477.1206N} {477, 1206}

\bibitem[\protect\citeauthoryear{{Nanayakkara} et~al.,}{{Nanayakkara} et~al.}{2022}]{Nanayakkara_2022}
{Nanayakkara} T.,  et~al., 2022, \mn@doi [arXiv e-prints] {10.48550/arXiv.2212.11638}, \href {https://ui.adsabs.harvard.edu/abs/2022arXiv221211638N} {p. arXiv:2212.11638}

\bibitem[\protect\citeauthoryear{{Nelson} et~al.,}{{Nelson} et~al.}{2018}]{Nelson_2018}
{Nelson} D.,  et~al., 2018, \mn@doi [\mnras] {10.1093/mnras/stx3040}, \href {https://ui.adsabs.harvard.edu/abs/2018MNRAS.475..624N} {475, 624}

\bibitem[\protect\citeauthoryear{{Nelson} et~al.,}{{Nelson} et~al.}{2019}]{Nelson_2019}
{Nelson} D.,  et~al., 2019, \mn@doi [Computational Astrophysics and Cosmology] {10.1186/s40668-019-0028-x}, \href {https://ui.adsabs.harvard.edu/abs/2019ComAC...6....2N} {6, 2}

\bibitem[\protect\citeauthoryear{{Oser}, {Ostriker}, {Naab}, {Johansson}  \& {Burkert}}{{Oser} et~al.}{2010}]{Oser_2010}
{Oser} L.,  {Ostriker} J.~P.,  {Naab} T.,  {Johansson} P.~H.,   {Burkert} A.,  2010, \mn@doi [\apj] {10.1088/0004-637X/725/2/2312}, \href {https://ui.adsabs.harvard.edu/abs/2010ApJ...725.2312O} {725, 2312}

\bibitem[\protect\citeauthoryear{{Park} et~al.,}{{Park} et~al.}{2022}]{Park_2022}
{Park} M.,  et~al., 2022, \mn@doi [\mnras] {10.1093/mnras/stac1773}, \href {https://ui.adsabs.harvard.edu/abs/2022MNRAS.515..213P} {515, 213}

\bibitem[\protect\citeauthoryear{{Park} et~al.,}{{Park} et~al.}{2023}]{Park_2023}
{Park} M.,  et~al., 2023, \mn@doi [\apj] {10.3847/1538-4357/acd54a}, \href {https://ui.adsabs.harvard.edu/abs/2023ApJ...953..119P} {953, 119}

\bibitem[\protect\citeauthoryear{Peng, Maiolino  \& Cochrane}{Peng et~al.}{2015}]{Peng_2015}
Peng Y.,  Maiolino R.,   Cochrane R.,  2015, \mn@doi [Nature] {10.1038/nature14439}, 521, 192–195

\bibitem[\protect\citeauthoryear{Pillepich et~al.,}{Pillepich et~al.}{2017}]{Pillepich_2017}
Pillepich A.,  et~al., 2017, \mn@doi [\mnras] {10.1093/mnras/stx2656}, 473, 4077–4106

\bibitem[\protect\citeauthoryear{{Pillepich} et~al.,}{{Pillepich} et~al.}{2018}]{Pillepich_2018}
{Pillepich} A.,  et~al., 2018, \mn@doi [\mnras] {10.1093/mnras/stx3112}, \href {https://ui.adsabs.harvard.edu/abs/2018MNRAS.475..648P} {475, 648}

\bibitem[\protect\citeauthoryear{{Quai}, {Hani}, {Ellison}, {Patton}  \& {Woo}}{{Quai} et~al.}{2021}]{Quai_2021}
{Quai} S.,  {Hani} M.~H.,  {Ellison} S.~L.,  {Patton} D.~R.,   {Woo} J.,  2021, \mn@doi [\mnras] {10.1093/mnras/stab988}, \href {https://ui.adsabs.harvard.edu/abs/2021MNRAS.504.1888Q} {504, 1888}

\bibitem[\protect\citeauthoryear{{Ramesh}, {Nelson}  \& {Pillepich}}{{Ramesh} et~al.}{2023}]{Ramesh_2023}
{Ramesh} R.,  {Nelson} D.,   {Pillepich} A.,  2023, \mn@doi [\mnras] {10.1093/mnras/stac3524}, \href {https://ui.adsabs.harvard.edu/abs/2023MNRAS.518.5754R} {518, 5754}

\bibitem[\protect\citeauthoryear{{Rees} \& {Ostriker}}{{Rees} \& {Ostriker}}{1977}]{Rees_Ostriker_1997}
{Rees} M.~J.,  {Ostriker} J.~P.,  1977, \mn@doi [\mnras] {10.1093/mnras/179.4.541}, \href {https://ui.adsabs.harvard.edu/abs/1977MNRAS.179..541R} {179, 541}

\bibitem[\protect\citeauthoryear{{Rennehan}, {Babul}, {Moa}  \& {Dav{\'e}}}{{Rennehan} et~al.}{2023}]{Rennehan_2023}
{Rennehan} D.,  {Babul} A.,  {Moa} B.,   {Dav{\'e}} R.,  2023, \mn@doi [arXiv e-prints] {10.48550/arXiv.2309.15898}, \href {https://ui.adsabs.harvard.edu/abs/2023arXiv230915898R} {p. arXiv:2309.15898}

\bibitem[\protect\citeauthoryear{Rodriguez-Gomez et~al.,}{Rodriguez-Gomez et~al.}{2015}]{Rodriguez_Gomez_2015}
Rodriguez-Gomez V.,  et~al., 2015, \mn@doi [\mnras] {10.1093/mnras/stv264}, 449, 49–64

\bibitem[\protect\citeauthoryear{{Rodr{\'\i}guez Montero}, {Dav{\'e}}, {Wild}, {Angl{\'e}s-Alc{\'a}zar}  \& {Narayanan}}{{Rodr{\'\i}guez Montero} et~al.}{2019}]{Curro_2019}
{Rodr{\'\i}guez Montero} F.,  {Dav{\'e}} R.,  {Wild} V.,  {Angl{\'e}s-Alc{\'a}zar} D.,   {Narayanan} D.,  2019, \mn@doi [\mnras] {10.1093/mnras/stz2580}, \href {https://ui.adsabs.harvard.edu/abs/2019MNRAS.490.2139R} {490, 2139}

\bibitem[\protect\citeauthoryear{{Rupke} \& {Veilleux}}{{Rupke} \& {Veilleux}}{2013}]{Rupke_2013}
{Rupke} D. S.~N.,  {Veilleux} S.,  2013, \mn@doi [\apj] {10.1088/0004-637X/768/1/75}, \href {https://ui.adsabs.harvard.edu/abs/2013ApJ...768...75R} {768, 75}

\bibitem[\protect\citeauthoryear{Santini et~al.,}{Santini et~al.}{2021}]{Santini_2021}
Santini P.,  et~al., 2021, \mn@doi [\aap] {10.1051/0004-6361/202039738}, 652, A30

\bibitem[\protect\citeauthoryear{{Schaye} et~al.,}{{Schaye} et~al.}{2015}]{Schaye_2015}
{Schaye} J.,  et~al., 2015, \mn@doi [\mnras] {10.1093/mnras/stu2058}, \href {https://ui.adsabs.harvard.edu/abs/2015MNRAS.446..521S} {446, 521}

\bibitem[\protect\citeauthoryear{{Schreiber} et~al.,}{{Schreiber} et~al.}{2018}]{Schreiber_2018}
{Schreiber} C.,  et~al., 2018, \mn@doi [\aap] {10.1051/0004-6361/201833070}, \href {https://ui.adsabs.harvard.edu/abs/2018A&A...618A..85S} {618, A85}

\bibitem[\protect\citeauthoryear{Schreiber et~al.,}{Schreiber et~al.}{2019}]{Schreiber_2019}
Schreiber N. M.~F.,  et~al., 2019, \mn@doi [\apj] {10.3847/1538-4357/ab0ca2}, 875, 21

\bibitem[\protect\citeauthoryear{{Sijacki}, {Springel}, {Di Matteo}  \& {Hernquist}}{{Sijacki} et~al.}{2007}]{Sijacki_2007}
{Sijacki} D.,  {Springel} V.,  {Di Matteo} T.,   {Hernquist} L.,  2007, \mn@doi [\mnras] {10.1111/j.1365-2966.2007.12153.x}, \href {https://ui.adsabs.harvard.edu/abs/2007MNRAS.380..877S} {380, 877}

\bibitem[\protect\citeauthoryear{Sijacki, Vogelsberger, Genel, Springel, Torrey, Snyder, Nelson  \& Hernquist}{Sijacki et~al.}{2015}]{Sijacki_2015}
Sijacki D.,  Vogelsberger M.,  Genel S.,  Springel V.,  Torrey P.,  Snyder G.~F.,  Nelson D.,   Hernquist L.,  2015, \mn@doi [\mnras] {10.1093/mnras/stv1340}, 452, 575–596

\bibitem[\protect\citeauthoryear{Smethurst, Lintott, Bamford, Hart, Kruk, Masters, Nichol  \& Simmons}{Smethurst et~al.}{2017}]{Smethurst_2017}
Smethurst R.~J.,  Lintott C.~J.,  Bamford S.~P.,  Hart R.~E.,  Kruk S.~J.,  Masters K.~L.,  Nichol R.~C.,   Simmons B.~D.,  2017, \mn@doi [\mnras] {10.1093/mnras/stx973}, 469, 3670–3687

\bibitem[\protect\citeauthoryear{{Soliman}, {Macci{\`o}}  \& {Blank}}{{Soliman} et~al.}{2023}]{Soliman_2023}
{Soliman} N.~H.,  {Macci{\`o}} A.~V.,   {Blank} M.,  2023, \mn@doi [\mnras] {10.1093/mnras/stad2295}, \href {https://ui.adsabs.harvard.edu/abs/2023MNRAS.525...12S} {525, 12}

\bibitem[\protect\citeauthoryear{{Somerville}, {Hopkins}, {Cox}, {Robertson}  \& {Hernquist}}{{Somerville} et~al.}{2008}]{Somerville_2008}
{Somerville} R.~S.,  {Hopkins} P.~F.,  {Cox} T.~J.,  {Robertson} B.~E.,   {Hernquist} L.,  2008, \mn@doi [\mnras] {10.1111/j.1365-2966.2008.13805.x}, \href {https://ui.adsabs.harvard.edu/abs/2008MNRAS.391..481S} {391, 481}

\bibitem[\protect\citeauthoryear{{Somerville} et~al.,}{{Somerville} et~al.}{2018}]{Somerville_2018}
{Somerville} R.~S.,  et~al., 2018, \mn@doi [\mnras] {10.1093/mnras/stx2040}, \href {https://ui.adsabs.harvard.edu/abs/2018MNRAS.473.2714S} {473, 2714}

\bibitem[\protect\citeauthoryear{{Springel}}{{Springel}}{2010}]{Springel_2010}
{Springel} V.,  2010, \mn@doi [\mnras] {10.1111/j.1365-2966.2009.15715.x}, \href {https://ui.adsabs.harvard.edu/abs/2010MNRAS.401..791S} {401, 791}

\bibitem[\protect\citeauthoryear{{Springel} \& {Hernquist}}{{Springel} \& {Hernquist}}{2003}]{Springel&Hernquist_2003}
{Springel} V.,  {Hernquist} L.,  2003, \mn@doi [\mnras] {10.1046/j.1365-8711.2003.06206.x}, \href {https://ui.adsabs.harvard.edu/abs/2003MNRAS.339..289S} {339, 289}

\bibitem[\protect\citeauthoryear{{Springel}, {White}, {Tormen}  \& {Kauffmann}}{{Springel} et~al.}{2001}]{Springel_2001}
{Springel} V.,  {White} S. D.~M.,  {Tormen} G.,   {Kauffmann} G.,  2001, \mn@doi [\mnras] {10.1046/j.1365-8711.2001.04912.x}, \href {https://ui.adsabs.harvard.edu/abs/2001MNRAS.328..726S} {328, 726}

\bibitem[\protect\citeauthoryear{{Springel}, {Di Matteo}  \& {Hernquist}}{{Springel} et~al.}{2005}]{Springel_2005}
{Springel} V.,  {Di Matteo} T.,   {Hernquist} L.,  2005, \mn@doi [\mnras] {10.1111/j.1365-2966.2005.09238.x}, \href {https://ui.adsabs.harvard.edu/abs/2005MNRAS.361..776S} {361, 776}

\bibitem[\protect\citeauthoryear{{Springel} et~al.,}{{Springel} et~al.}{2018}]{Springel_2018}
{Springel} V.,  et~al., 2018, \mn@doi [\mnras] {10.1093/mnras/stx3304}, \href {https://ui.adsabs.harvard.edu/abs/2018MNRAS.475..676S} {475, 676}

\bibitem[\protect\citeauthoryear{Sturm et~al.,}{Sturm et~al.}{2011}]{Sturm_2011}
Sturm E.,  et~al., 2011, \mn@doi [\apj] {10.1088/2041-8205/733/1/l16}, 733, L16

\bibitem[\protect\citeauthoryear{{Talbot}, {Sijacki}  \& {Bourne}}{{Talbot} et~al.}{2022}]{Talbot_2022}
{Talbot} R.~Y.,  {Sijacki} D.,   {Bourne} M.~A.,  2022, \mn@doi [\mnras] {10.1093/mnras/stac1566}, \href {https://ui.adsabs.harvard.edu/abs/2022MNRAS.514.4535T} {514, 4535}

\bibitem[\protect\citeauthoryear{{Talbot}, {Sijacki}  \& {Bourne}}{{Talbot} et~al.}{2024}]{Talbot_2024}
{Talbot} R.~Y.,  {Sijacki} D.,   {Bourne} M.~A.,  2024, \mn@doi [\mnras] {10.1093/mnras/stae392}, \href {https://ui.adsabs.harvard.edu/abs/2024MNRAS.528.5432T} {528, 5432}

\bibitem[\protect\citeauthoryear{{Terrazas}, {Bell}, {Woo}  \& {Henriques}}{{Terrazas} et~al.}{2017}]{Terrazas_2017}
{Terrazas} B.~A.,  {Bell} E.~F.,  {Woo} J.,   {Henriques} B. M.~B.,  2017, \mn@doi [\apj] {10.3847/1538-4357/aa7d07}, \href {https://ui.adsabs.harvard.edu/abs/2017ApJ...844..170T} {844, 170}

\bibitem[\protect\citeauthoryear{{Terrazas} et~al.,}{{Terrazas} et~al.}{2020}]{Terrazas_2020}
{Terrazas} B.~A.,  et~al., 2020, \mn@doi [\mnras] {10.1093/mnras/staa374}, \href {https://ui.adsabs.harvard.edu/abs/2020MNRAS.493.1888T} {493, 1888}

\bibitem[\protect\citeauthoryear{{Thompson}, {Fabian}, {Quataert}  \& {Murray}}{{Thompson} et~al.}{2015}]{Thompson_2015}
{Thompson} T.~A.,  {Fabian} A.~C.,  {Quataert} E.,   {Murray} N.,  2015, \mn@doi [\mnras] {10.1093/mnras/stv246}, \href {https://ui.adsabs.harvard.edu/abs/2015MNRAS.449..147T} {449, 147}

\bibitem[\protect\citeauthoryear{Torrey et~al.,}{Torrey et~al.}{2019}]{Torrey_2019}
Torrey P.,  et~al., 2019, \mn@doi [\mnras] {10.1093/mnras/stz243}

\bibitem[\protect\citeauthoryear{{Trayford}, {Theuns}, {Bower}, {Crain}, {Lagos}, {Schaller}  \& {Schaye}}{{Trayford} et~al.}{2016}]{Trayford_2016}
{Trayford} J.~W.,  {Theuns} T.,  {Bower} R.~G.,  {Crain} R.~A.,  {Lagos} C. d.~P.,  {Schaller} M.,   {Schaye} J.,  2016, \mn@doi [\mnras] {10.1093/mnras/stw1230}, \href {https://ui.adsabs.harvard.edu/abs/2016MNRAS.460.3925T} {460, 3925}

\bibitem[\protect\citeauthoryear{{Tremmel}, {Karcher}, {Governato}, {Volonteri}, {Quinn}, {Pontzen}, {Anderson}  \& {Bellovary}}{{Tremmel} et~al.}{2017}]{Tremmel_2017}
{Tremmel} M.,  {Karcher} M.,  {Governato} F.,  {Volonteri} M.,  {Quinn} T.~R.,  {Pontzen} A.,  {Anderson} L.,   {Bellovary} J.,  2017, \mn@doi [\mnras] {10.1093/mnras/stx1160}, \href {https://ui.adsabs.harvard.edu/abs/2017MNRAS.470.1121T} {470, 1121}

\bibitem[\protect\citeauthoryear{{Valentino} et~al.,}{{Valentino} et~al.}{2020}]{Valentino_2020}
{Valentino} F.,  et~al., 2020, \mn@doi [\apj] {10.3847/1538-4357/ab64dc}, \href {https://ui.adsabs.harvard.edu/abs/2020ApJ...889...93V} {889, 93}

\bibitem[\protect\citeauthoryear{{Valentino} et~al.,}{{Valentino} et~al.}{2023}]{Valentino_2023}
{Valentino} F.,  et~al., 2023, \mn@doi [\apj] {10.3847/1538-4357/acbefa}, \href {https://ui.adsabs.harvard.edu/abs/2023ApJ...947...20V} {947, 20}

\bibitem[\protect\citeauthoryear{{Vogelsberger}, {Genel}, {Sijacki}, {Torrey}, {Springel}  \& {Hernquist}}{{Vogelsberger} et~al.}{2013}]{Vogelsberger_2013}
{Vogelsberger} M.,  {Genel} S.,  {Sijacki} D.,  {Torrey} P.,  {Springel} V.,   {Hernquist} L.,  2013, \mn@doi [\mnras] {10.1093/mnras/stt1789}, \href {https://ui.adsabs.harvard.edu/abs/2013MNRAS.436.3031V} {436, 3031}

\bibitem[\protect\citeauthoryear{{Vogelsberger} et~al.,}{{Vogelsberger} et~al.}{2014}]{Vogelsberger_2014}
{Vogelsberger} M.,  et~al., 2014, \mn@doi [\mnras] {10.1093/mnras/stu1536}, \href {https://ui.adsabs.harvard.edu/abs/2014MNRAS.444.1518V} {444, 1518}

\bibitem[\protect\citeauthoryear{Vogelsberger et~al.,}{Vogelsberger et~al.}{2018}]{Vogelsberger_2018}
Vogelsberger M.,  et~al., 2018, \mn@doi [\mnras] {10.1093/mnras/stx2955}, 474, 2073–2093

\bibitem[\protect\citeauthoryear{{Wagner} \& {Bicknell}}{{Wagner} \& {Bicknell}}{2011}]{Wagner_Bicknell_2011}
{Wagner} A.~Y.,  {Bicknell} G.~V.,  2011, \mn@doi [\apj] {10.1088/0004-637X/728/1/29}, \href {https://ui.adsabs.harvard.edu/abs/2011ApJ...728...29W} {728, 29}

\bibitem[\protect\citeauthoryear{{Wagner}, {Bicknell}  \& {Umemura}}{{Wagner} et~al.}{2012}]{Wagner_2012}
{Wagner} A.~Y.,  {Bicknell} G.~V.,   {Umemura} M.,  2012, \mn@doi [\apj] {10.1088/0004-637X/757/2/136}, \href {https://ui.adsabs.harvard.edu/abs/2012ApJ...757..136W} {757, 136}

\bibitem[\protect\citeauthoryear{{Weaver} et~al.,}{{Weaver} et~al.}{2022}]{Weaver2022}
{Weaver} J.~R.,  et~al., 2022, \mn@doi [arXiv e-prints] {10.48550/arXiv.2212.02512}, \href {https://ui.adsabs.harvard.edu/abs/2022arXiv221202512W} {p. arXiv:2212.02512}

\bibitem[\protect\citeauthoryear{{Weinberger} et~al.,}{{Weinberger} et~al.}{2017}]{Weinberger_2016}
{Weinberger} R.,  et~al., 2017, \mn@doi [\mnras] {10.1093/mnras/stw2944}, \href {https://ui.adsabs.harvard.edu/abs/2017MNRAS.465.3291W} {465, 3291}

\bibitem[\protect\citeauthoryear{{Weinberger} et~al.,}{{Weinberger} et~al.}{2018}]{Weinberger_2018}
{Weinberger} R.,  et~al., 2018, \mn@doi [\mnras] {10.1093/mnras/sty1733}, \href {https://ui.adsabs.harvard.edu/abs/2018MNRAS.479.4056W} {479, 4056}

\bibitem[\protect\citeauthoryear{{Westmoquette}, {Clements}, {Bendo}  \& {Khan}}{{Westmoquette} et~al.}{2012}]{Westmoquette_2012}
{Westmoquette} M.~S.,  {Clements} D.~L.,  {Bendo} G.~J.,   {Khan} S.~A.,  2012, \mn@doi [\mnras] {10.1111/j.1365-2966.2012.21214.x}, \href {https://ui.adsabs.harvard.edu/abs/2012MNRAS.424..416W} {424, 416}

\bibitem[\protect\citeauthoryear{{Wetzel}, {Tinker}, {Conroy}  \& {van den Bosch}}{{Wetzel} et~al.}{2013}]{Wetzel_2013}
{Wetzel} A.~R.,  {Tinker} J.~L.,  {Conroy} C.,   {van den Bosch} F.~C.,  2013, \mn@doi [\mnras] {10.1093/mnras/stt469}, \href {https://ui.adsabs.harvard.edu/abs/2013MNRAS.432..336W} {432, 336}

\bibitem[\protect\citeauthoryear{{Yang} \& {Reynolds}}{{Yang} \& {Reynolds}}{2016}]{Yang_2016}
{Yang} H. Y.~K.,  {Reynolds} C.~S.,  2016, \mn@doi [\apj] {10.3847/0004-637X/829/2/90}, \href {https://ui.adsabs.harvard.edu/abs/2016ApJ...829...90Y} {829, 90}

\bibitem[\protect\citeauthoryear{{Zinger} et~al.,}{{Zinger} et~al.}{2020}]{Zinger_2020}
{Zinger} E.,  et~al., 2020, \mn@doi [\mnras] {10.1093/mnras/staa2607}, \href {https://ui.adsabs.harvard.edu/abs/2020MNRAS.499..768Z} {499, 768}

\makeatother
\end{thebibliography}

\appendix

\bsp	          
\label{lastpage}
\end{document}